%% file: sample631.tex
\documentclass[twocolumn,trackchanges,astrosymb]{aastex631}

\newcommand\feh{\rm{[Fe/H]}}
\newcommand\teff{\rm{T_{eff}}}
\newcommand\logg{\rm{\log(g)}}
\newcommand\loglbol{\rm{\log(L/L_{\odot})}}
\newcommand\ali{\rm{A(Li)}}
\newcommand\detfrac{$\frac{N_{Detections}}{N_{Detections} + N_{Upper Limits}}$}
\newcommand\msun{\mathrm{M_{\odot}}}
\newcommand\mscaled{\mathrm{M_{scaled}}}

\usepackage{float}
\usepackage{multirow}
\usepackage{booktabs}
\usepackage{amsmath}
\usepackage{tabularx} % Add in the preamble
\usepackage{graphicx}
\usepackage{subcaption}

\begin{document}

\title{Let Sleeping Stars Li: \\ Mapping Turnoff Lithium, the Lithium Dip,  \\ \& Testing Standard Model Predictions in GALAH DR4 Subgiants}

\author[0009-0008-8819-0481]{Vincent A. Smedile}
\affiliation{The Ohio State University, Department of Astronomy, 140 W. 18th Avenue, Columbus, OH 43210, USA}
\affiliation{Center for Cosmology \& AstroParticle Physics, 191 W. Woodruff Avenue, Columbus, OH 43210, USA}

\author[0000-0002-7549-7766]{Marc H. Pinsonneault}
\affiliation{The Ohio State University, Department of Astronomy, 140 W. 18th Avenue, Columbus, OH 43210, USA}
\affiliation{Center for Cosmology \& AstroParticle Physics, 191 W. Woodruff Avenue, Columbus, OH 43210, USA}

\begin{abstract}
Lithium is an excellent tracer of stellar mixing. We reconstruct the end of the main sequence lithium depletion pattern using the GALAH DR4 3D-NLTE lithium dataset for 22,442 stars, complemented by masses and subgiant lithium dilution factors inferred from YREC stellar models. We recover the two canonical features seen in clusters: a cool plateau ($\mathrm{M_{\feh_{\odot}}} \sim 1.2 \mathrm{M_\odot}$) with a median $\ali = 2.42$, and a steep plunge into the lithium dip ($\mathrm{M_{\feh_{\odot}}} \sim 1.35 \mathrm{M_\odot}$) dominated by upper limits at $\ali < 1.5$. Above the dip, our data reveal a striking new pattern with a wide range of abundances. We argue this reflects severe rotationally induced mixing in most stars, with lithium preserved in a minority where gravitational settling suppresses mixing. Lithium depletion with intrinsic dispersion is universal across the sample. Predicted and observed dilution trends agree well below the dip. For higher-mass stars, dilution corrections agree with standard models, but our upper envelope lies below open cluster values. We attribute this to a shallower lithium preservation zone and further evidence for gravitational settling and radiative levitation. We detect lithium in $\approx25\%$ of dip stars, whereas clusters show none; neither blurring of a narrow mass feature nor binary contamination fully explains this. The cool plateau is 0.8 dex below the solar meteoritic value across a wide range of $\feh$. After accounting for pre-main sequence depletion, the plateau is elevated at $\geq\feh_{\odot}$, possibly evidence for lithium production.

\end{abstract} \label{abstract}


\section{Introduction}
Lithium (lithium) is a fragile but powerful tracer of stellar physics. Lithium burns at $\teff \gtrsim 2.5\times10^{6}$ K, a threshold reached in the outer $0.02\msun$ of most stars. Lithium consequently serves as a sensitive diagnostic of internal mixing \citep{marcannualreview, charbonnelreview}. 

Lithium also has an interesting, and complex, origin. It is produced in the Big Bang \citep{bosegaard1985bbn}, but there are also significant and post-primordial Galactic sources \citep{cameronfowler,rgbcameronfowler,woosley,novaeli,marcannualreview,zahn2005spiteplat,borisov2024}. However, the observed stellar lithium abundance ($^{7}$[Li/H]${\rm_{obs}} \times 10^{-10} = 1.58^{+0.35}_{-0.28}$) is significantly lower than the BBN value ($^{7}$[Li/H]$_{\rm BBN} \times 10^{-10} = 5.627\pm0.259$) \citep{bbnlilatest}, in low-metallicity stars \citep{spiteandspite,thorburncoolplataeu}. Stellar models can reconcile the two with mild envelope mixing, which carries lithium to layers where it can be destroyed \citep{marcpapercosmoli,marc1992halostars,zahn1992,marc2001convectivezonefgk,zahn2005spiteplat, borisov2024}. The degree of lithium depletion in stars can therefore have broader astrophysical consequences.

In classical stellar models, convection is the sole mixing process. Prior to the main sequence, low-mass stars are predicted to have deep and hot convection zones, which can induce significant depletion. This pre-main sequence depletion increases strongly at lower mass \citep{bodenheimerpmsdepletin, proffittmichaud1989,dantona1994,marcannualreview,bildsten1997,ushomirsky1998}. On the main sequence, however, convection zones are too thin to burn lithium for solar-mass stars and above \citep{weyman_sears_convectivedepth, marcannualreview}. By the 1970s, early open-cluster observations already showed clear mass-dependent MS lithium depletion and a dispersion at fixed mass not seen in other species \citep{herbig1965,herbigsubgiants,zapaladepletioninfstars}. These early results were confirmed by a burst of later work in numerous cluster studies, for example in the Pleiades \citep{soderblom1995} and the Hyades \citep{thorburnhyades}, with a clear summary by \citet{sestrand2005}. More recently, there has been interesting work focusing on early F and late A stars \citep{ngc7789lidip, 2025arXiv250704266S}. 

Rotationally induced mixing is an appealing explanation \citep{marcrotmix1989,zahn1992}, and has been explored extensively in the context of clusters \citep{chaboyerrimm67gkstars,marcpapercosmoli,marcannualreview,borisov2024,charbonnelreview}. Still, a direct causal link remains elusive and remains a subject of discussion \citep{Theveninnotrotmix,reggianilibe,navarronotrotmix}. At fixed mass, stars lose memory of their initial rotation rates, so a direct link between observed rotation and mixing is neither seen nor expected.

Lithium in stars is also sensitive to other physical processes. Gravitational settling and thermal diffusion can cause heavy species to sink relative to light ones \citep{burgers1969,thoulgravdiffu1994}, important for solar models where it causes a $\sim10\%$ change in lithium \citep{bahcallandmarc,marc1992halostars,Soutom67}. Radiation pressure on large ions can cause radiative levitation, concentrating such species in near-surface layers. These effects are especially relevant in early F and late A stars where they can also interact with rotational mixing \citep{Michaud1970,diffusionpaper,abt1979,Abt1995,vauclair2005,Abt_2017}. A particular example is in slowly rotating A stars; the interactions between levitation and internal magnetic-field effects suppress rotational mixing while enhancing microscopic diffusion of specific species \citep{diffusionpaper,mieridonalcircdiffusion, vauclair2005,borisov2024}, resulting in chemical peculiarity in all such stars \citep{Michaud1970,abt1979,Abt1995,Abt_2017}. 

Stars also sharply transition from slow to rapid rotators around the mid-F region, known as the Kraft Break \citep{1965kraftbreak2,kratbreak1967}. This transition is related to the transition from deep to shallow surface convection zones and the cessation of magnetized winds. The transition between the cool and hot star regimes is also characterized by severe lithium depletion \citep{bosegaard1986,bosegaard1987}, known as the lithium dip. We know this to be a region of true lithium depletion and destruction based on subgiant data \citep{bosegaard1986,bosegaard1987, Balachanrdanm67,marcrotmix1989,stillsdeliyannis2000,2019deliyannis}. The Li dip discovery papers also reported a steep rise in Li for hotter stars (thus the "dip"). While initially thought to be a product of diffusion processes \citep{michauddiffusionfstars}, rotation-based processes have become increasingly favored \citep{zahn1992, talonandcharblidip,ngc188rotmixlidip}. The full breadth of causes remains an active subject (see \citealt{marcannualreview,charbonnelreview}, \& references therein).

Historically, it has been difficult to measure lithium for stars hotter than the dip. The main spectroscopic feature, the lithium I 6707 Å resonance doublet, becomes difficult to detect in hot stars \citep{randich2021}. Selection effects are also a concern. Because it is more difficult to detect lithium in rapid rotators than in slow ones, the observed sample will be biased. The turnoff region of open clusters also contains large numbers of blue stragglers, and distinguishing binaries from single stars close to the turnoff is non-trivial.

Despite these limitations, recent work in open clusters \citep{stillsdeliyannis2000,2019deliyannis,gao_m67paper,ngc188lidip, ngc7789lidip} has increased sample sizes significantly for hot stars. These studies observed a clear population of stars with high abundances, similar to the populations seen in the early studies. In these larger samples, there is a majority of stars with detections at high levels and significant dispersion.

Star clusters have therefore proven to be a rich source of data about lithium depletion. However, these well-studied open clusters are close to solar metallicity and are much smaller than large modern survey samples. Fortunately, large field catalogs have recently become available. \cite{gao2020} used the GALactic Archaeology with Hermes (GALAH) survey Second Data Release \citep{galahdr1, galahdr2, galahdr3, galahdr4} to produce the largest high-resolution analyses of stellar lithium, comprising $\sim100,000$ stars. They observed similar abundance trends in $\ali$-$\feh$ between the stars above and below the lithium dip, and noted a small offset of $\ali\sim0.4$ dex. However, their work was largely qualitative; it did not examine the upper limits or attempt to correct for mixing and dilution effects in cool stars.

Our central goal is to infer the lithium depletion pattern at the end of the main sequence as a function of mass and metallicity. Subgiants are ideal targets: their HR diagram positions map well onto mass and age, and the subgiant branch is well-populated across a wide range of masses and metallicities. Intermediate-mass subgiants are also cooler and slower-rotating than their MS counterparts, letting us study the full lithium pattern for more massive stars than are directly accessible on the main sequence.

The published data at face value, however, have limitations. A large fraction are upper limits, warranting proper statistical treatment. As stars evolve from the MS to the RGB, deep surface convection zones dilute lithium for $\teff \lesssim 5800$~K in a mass- and metallicity-dependent manner, even in standard models. This First Dredge-Up (FDU) event \citep{iben1974review,kippenhanfdu} grows severe toward the RGB and must be accounted for. Lithium also burns near the RGB base in low-mass, high-metallicity models \citep{kailimarc2025}, and model-based masses face strong systematics there too. HR diagram positions near the field turnoff or RGB may further overlap with photometric binaries and other contaminants, respectively, so identifying such features is essential wherever our mass and age estimates could go wrong. Motivated by this, we re-analyze GALAH DR4 using our own stellar models for lithium dilution and mass inference.

The paper proceeds as follows: Section~\ref{Data and Methdology} covers data, models, and mass determinations; Section~\ref{Subgiant catalog} presents sample selection, tests standard-model FDU predictions, defines catalog bounds. Section~\ref{lithium pattern} calculates the lithium dip location, and analyzes the turnoff pattern, and relates the findings of the prior sections to the older GALAH field lithium study by \citet{gao2020}; Section~\ref{lirichgiants} uses turnoff abundances to predict future RGB depletion; Section~\ref{discussion} summarizes findings and conclusions.

\section{Data \& Theoretical Models} \label{Data and Methdology}

\subsection{GALAH: Base Dataset for Stellar Parameters and Lithium}
Our base dataset is the GALAH DR4 3D NLTE Lithium Abundance Value-Added Catalogs (VACs) \citep{wang2024} -- providing us with $\loglbol$, $\teff$, and abundances. GALAH DR4 is a large survey focused on providing spectra, stellar properties, and chemical abundances of up to 32 elements for 1,085,520 field stars across most evolutionary states. These include stellar properties derived from a combination of GALAH and Gaia astrometry. There are also several VACs with additional information and built-in cross-matching between GALAH DR4, Gaia DR3, and 2MASS. 

\subsection{GALAH DR4 Initial Sampling} \label{Dataset Reduction}
We first remove stars with \texttt{S/N} $\leq $30, stars with poor abundance fits (\texttt{flag$\_$ALi} $<$ 2, \texttt{flag$\_$X$\_$Fe} $<$ 0, and \texttt{flag$\_$fe$\_$h} $<$ 0), and stars with poor pipeline fits (GALAH DR4 \texttt{Flag\_sp} $>$0). 

Highly extincted stars can have inaccurate HR diagram positions that can also bias derived masses and luminosities. To address this, \citet{galahdr4} apply star-by-star bolometric corrections of \citet{CasagrandeVandenBerg} at $E(B-V) = 0$ and post-correct the resulting bolometric magnitudes by subtracting the extinction $A_{K_{s}}$ directly. Extinction in $K_{s}$ is itself estimated star-by-star using the Rayleigh--Jeans Color Excess (RJCE) method, $A_{K_{s}} = 0.917\,(H - W_{2} - 0.08)$ \citep{majewski_rjce_extinct}, from 2MASS $H$ and WISE $W_{2}$ photometry. The pipeline further validates these estimates against $A_{K_{s}} \approx 0.36 \cdot E(B-V)$ \citep{cardelliextinction}, using $E(B-V)$ from the \citet{schlegeldustmap} dust map. Although the \citet{schlegeldustmap} map represents an upper limit on the extinction toward any individual star at a finite distance, GALAH is restricted to $|b| \geq 10^{\circ}$, where mean extinctions are modest \citep{galahdr3, galahdr4}. As an independent check, we cross-matched our sample star-by-star against the Gaia 3D dust map \citep{greenbaystar}. For our sample, we find the median absolute deviation (MAD) between the measurements to be $\text{MAD}_{\Delta E(B-V)} = 0.025$\, indicating extinction is already well-constrained.

\subsection{Gaia XPBoost: Parameter Quality Checks}
We employ the Gaia XPBoost catalog \citep{gaiaxgboost} to assess random and systematic errors, as the stellar properties in GALAH DR4 employ Gaia parallax-corrected measurements.

\subsection{Catalogs for Identifying Contaminants and Binaries}
Contaminants from different evolutionary phases can bias our mass measurements. Binary companions, for instance, contribute light that biases mass estimates based on luminosity \citep{berger2020binary2,berger2020binary1}. This is particularly severe near the main sequence, where subgiants can be close in luminosity to binary systems with two main sequence stars. Conversely, binaries are less of a factor in cool subgiants, because their HR diagram position does not overlap with that of a main sequence binary. However, a near-turnoff main sequence companion of similar mass could still affect the luminosity. We employ the Million Star Main-Sequence (MSMS) Binary Catalog \citep{msmsbinarycat}, the 400K Subgiant Catalog \citep{natafcatalog}, GALAH FGK Binaries VAC \citep{galahfgkbinary}, the Gaia DR3 Non-Single Stars (NSS) catalog \citep{gaiansscat}, and the Gaia DR3 RV Variability Dataset \citep{gaiavari_all} (the latter two catalogs employ Gaia Parallax-corrected temperatures, luminosities, and RVs) to identify binary contamination near the Main Sequence Turnoff region. In addition, the high-mass subgiants may intermingle with the lower-mass end of the core helium-burning stars (red clump). We therefore utilize the APOKASC-3 \citep{apokasc3} red clump (RC) sample to disentangle the true massive subgiants from these RC stars.

\subsection{Stellar Models and Isochrones}
We have computed our own stellar model grid for inferring mass, lithium dilution from the first dredge-up, and dilution relative to the turnoff. Constraining the complex field lithium pattern requires precise and accurate masses, and it is important to use self-consistent masses and dredge-up corrections \citep{marcannualreview,balachandran1990,sestrand2005}. We utilize the YREC stellar evolution code \citep{yreccoderelease}. To assess random and systematic errors, we compare our calculations with the MIST \citep{mist} and PARSEC models \citep{parsec1}. PARSEC is particularly important because it was used to infer stellar properties in GALAH. 
\subsubsection{The Theoretical Model Suite}
We computed a suite of stellar models spanning $0.80 \leq \mathrm{M/M_{\odot}} \leq 2.80$ and $-0.40 \lesssim \feh \lesssim 0.40$ to determine subgiant masses, calculate lithium depletion, isolate the subgiant population, and identify the domains of the first dredge-up dilution and lithium burning. These models cover the mass and metallicity range of a majority of the GALAH DR4 stars \citep{galahdr4}. Our models were produced using the Yale Rotating Evolution Code (YREC; \citealt{demarque1964, marcrotmix1989, yrecpaper, vansadersyrec2012, yreccoderelease}). We adopted a convective core overshoot of 0.2 pressure scale heights. Opacities were taken from \cite{badnellopacities} with the \cite{grevessesauval1998} heavy-element mixture. The equation of state followed \cite{rogersnayfonov}. Nuclear reaction rates were from \citet{adelberger2011nuclearcrosssec}, and the outer boundary conditions at $T = T_{\mathrm{eff}}$ were derived from \cite{kuruzcuz1997atmos} model atmospheres. Our solar models were calibrated to reproduce $L_{\odot} = 3.8275 \times 10^{33} \mathrm{erg \cdot s^{-1}}$,  $R_{\odot} = 6.957 \times 10^{10} \mathrm{cm}$, at and age of $\text{4.568 Gyr}$ \citep{prisasolarmetrics}. \cite{grevessesauval1998} adopted $Y = 0.248$, $Z = 0.01649$, and $Z/X = 0.02292$ for the solar mixture. Our mixing length $\alpha$, hydrogen, helium, and metal abundances were inferred with a solar model calibration, yielding a well-posed: $\alpha = 1.93271$, $Y = 0.248$, $Z = 0.01649$, and $Z/X = 0.0231$.

Our grid consisted of standard models, neglecting both microscopic diffusion and rotationally induced mixing. We did not attempt to model MS lithium depletion, as the goal of this project is to infer rather than to predict the end of MS lithium depletion pattern. However, we do need to correct the observed abundances of cool subgiants for dilution due to the development of a deep surface convection zone. We therefore measured the logarithmic depletion of lithium relative to the turnoff value, rather than doing an absolute calculation. We inferred this standard model dilution and destruction factor by interpolating across a grid of model tracks. This procedure is similar to that of \cite{chanamemarc}.

\subsubsection{Mass Interpolation} 
We compute masses from our YREC grid with the aid of the \texttt{Kiauhoku} code \citep{kiauhoku}. Based on the methods of \citet{mist0}, \texttt{Kiauhoku} converts evolutionary models into equivalent evolutionary points, forms a multi-index grid, and interpolates stellar parameters between models. Here, we interpolate mass as a function of \feh, luminosity, and $\teff$. We only inferred masses for stars on the subgiant branch (SGB), and defined the start of the SGB at a central hydrogen abundance of $\mathrm{X_{Cen}} \leq 10^{-4}$.

\subsection{Open Cluster Data}
The GALAH data give us a snapshot at the end of the MS lifetime, which we can compare with the rich cluster literature. Lithium abundance on the main sequence depends on age, so we use the Pleiades as a metric for younger clusters and the Hyades as a representative of intermediate-age clusters. For older clusters, NGC 188 and M67 serve as our reference, with the former also containing numerous subgiants. Additionally, recent detailed NGC 7789 observations provide useful comparisons for the lithium dip, the Kraft break, and the stars above the lithium dip. Our cluster ages, $\feh$, and sources are summarized in Table \ref{tab:clusters}. 
\begin{table}
\centering
\caption{Open clusters used for lithium abundance comparisons, ordered by age.}
\label{tab:clusters}
\begin{tabular}{lcc}
\toprule
Cluster & Age (Gyr) & [Fe/H] \\
\midrule
Pleiades    & $0.125$ & $+0.03 \pm 0.05$ \\
Hyades      & $0.650$ & $+0.13 \pm 0.05$ \\
NGC 7789    & $1.46$ & $-0.13 \pm 0.05$ \\
M67         & $4.0$ & $-0.01 \pm 0.05$ \\
NGC 188     & $6.3$ & $+0.064 \pm 0.016$ \\
\bottomrule
\end{tabular}
\par\smallskip
{\footnotesize \textbf{Notes.} Ages and metallicities for the Hyades, NGC 7789,
Pleiades, M67, and NGC 188 are drawn from \citet{Balachanrdanm67},
\citet{ngc7789lidip}, \citet{pleiadesbouvier}, \citet{paceclustm67}, and \citet{ngc188rotmixlidip}
respectively.}
\end{table}

\section{Subgiant \& Turnoff lithium Catalog} \label{Subgiant catalog}
\begin{figure}
\centering
\includegraphics[width=1\linewidth]{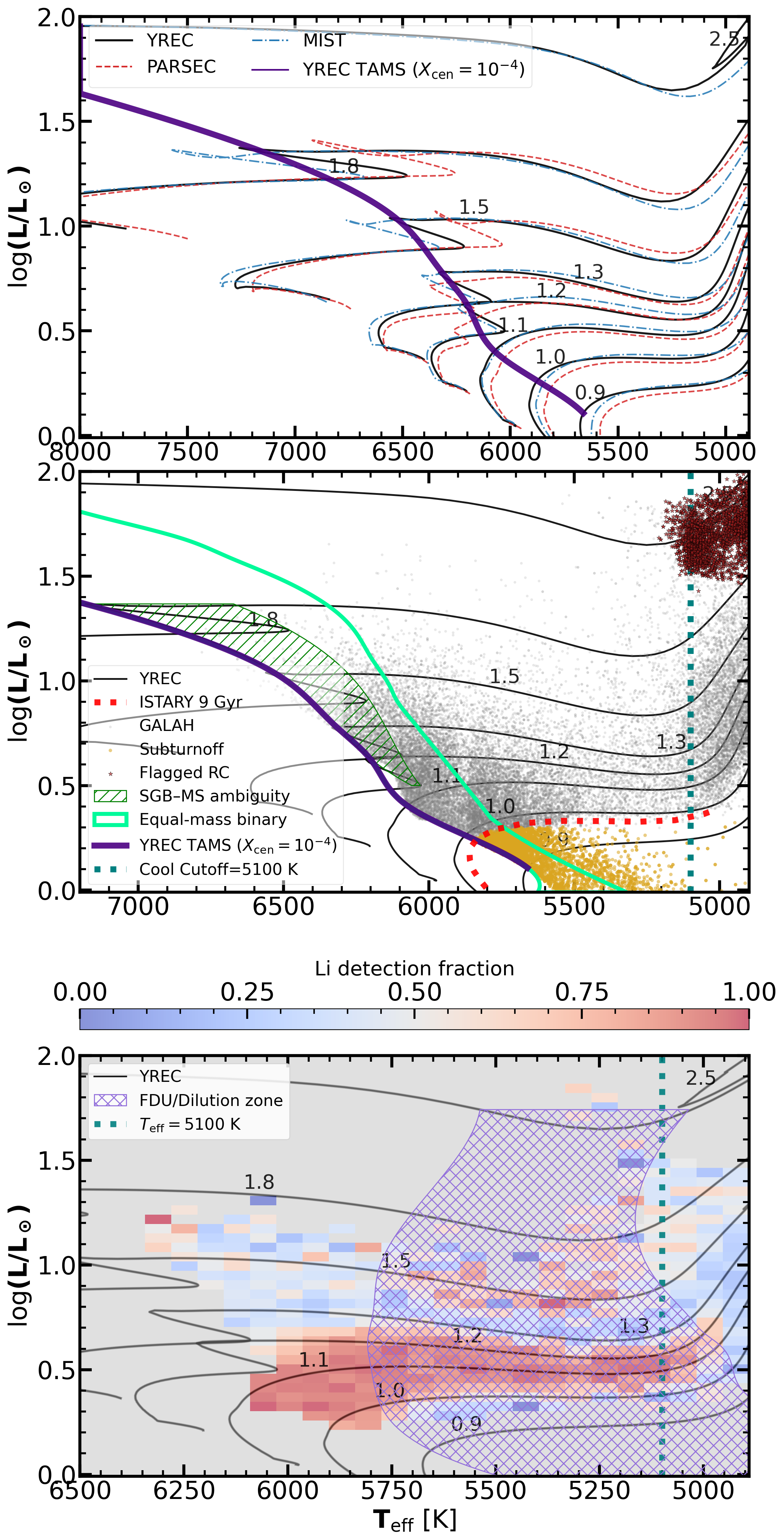}
\caption{Overview of the initial GALAH sample and theoretical tracks. \textbf{Top}: YREC models compared to those of MIST and PARSEC. The purple line marks the YREC-defined SGB start. \textbf{Middle}: YREC tracks over the initial GALAH sample, highlighting the SGB--MS ambiguous domain, the subturnoff, and the red clump. The bright teal line marks the expected equal-mass binary domain, $0.3$ dex in $\loglbol$ above the MS. \textbf{Bottom}: We define the onset of the FDU as occurring when lithium dilution exceeds 0.01 dex below the turnoff value. We also show the onset of severe depletion, which we define as the point where lithium dilution exceeds 1.5 dex below the turnoff value. We overlay the subgiant dilution factors. The hatched region spans from the onset of the first dredge-up to 1.5\,dex of dilution.}
\label{fig:catalog_sample_compare}
\end{figure}

Inferring subgiant masses from HR diagram position requires a domain where the mass inference is reliable and the systematic model errors are understood. For subgiants, it is especially important to remove photometric binaries and contaminants. Calculating the post-main sequence lithium depletion pattern also requires accounting for the development of deep surface convection zones, which can dilute or even destroy lithium. Because these depletion factors can be large, it is important to test whether the trends predicted in the models are seen in the data. Finally, a large fraction of the sample consists of upper limits. We don't wish to leave these out; therefore, their careful interpretation is required. We lay out the major model features and global sample properties in Figure~\ref{fig:catalog_sample_compare}.

We show different theoretical models in the top panel. Our subgiant models (solid black lines) are compared with PARSEC \citep{parsec1} and MIST \citep{mist} (dashed lines). It is encouraging that there is good agreement between the models in the mass-luminosity relationship. There is some difference in where the models define the onset of the subgiant branch, which is most obvious in the end of MS ``hooks" of more massive models. These offsets will translate to model-dependent differences in the subgiant domain in the HR diagram. We thus use the differences between our masses and the PARSEC-based ones from GALAH as both a measure of systematic uncertainties in mass estimates and a limit on the domain where we have reliable masses. We note that using PARSEC or MIST would have yielded a cooler or hotter turnoff boundary, respectively. We also compare GALAH data with independent measurements of abundance and $\teff$ to establish systematic uncertainties in those variables.

In the middle panel, we show the population effects that we need to consider. The purple line is our defined start of the subgiant branch. The dark green hatched area contains a mixture of MS and SGB stars, which we must remove because the MS stars will be far more numerous \citep{stellarmultiplicity}. We overlay the APOKASC-3 \citep{apokasc3} red clump (RC) sample to remove the domain where rare high-mass subgiants are intermingled with these more numerous stars. 
\begin{figure*}
\centering
\includegraphics[width=1\linewidth]{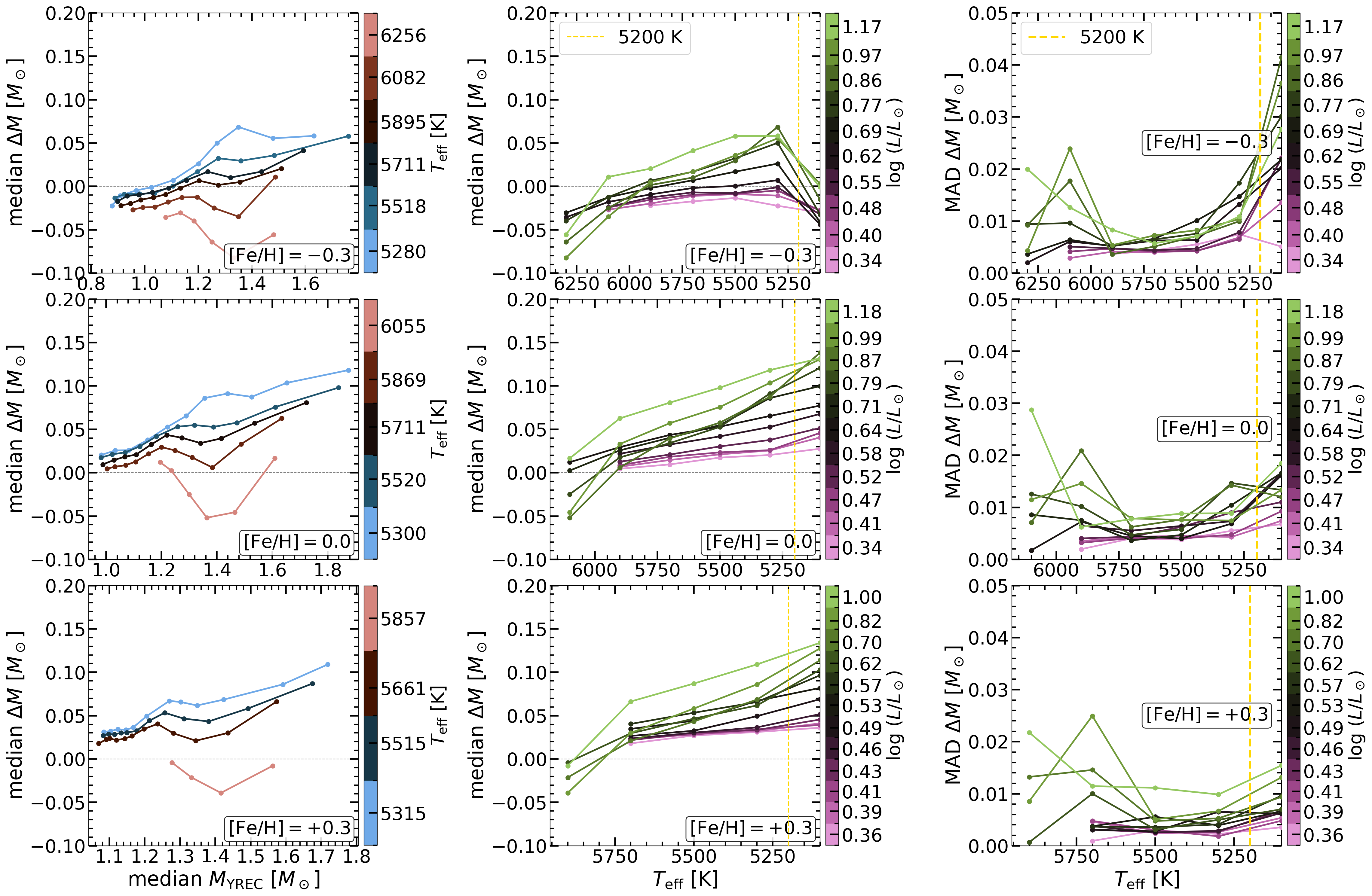}
\caption{An overview of the systematic offsets between the GALAH and YREC masses. The two parameters, being largely theoretically interpolated, inherit the systematics of their parent track sets (YREC and PARSEC). The rows represent $\feh \mathrm{ -0.3, 0.0, and +0.3}$. The \textbf{left column} plots mass differences as a function of YREC mass for vertical slices in $\teff$. The \textbf{ middle and right columns} show trends as a function of $\teff$ in slices at fixed L, while the right column shows the median absolute deviation (MAD) for the same slices.}
\label{fig:systemic_compare}
\end{figure*}

There are three other important population features: the finite age of the galaxy, main sequence binaries that overlap with the subgiant domain, and the impact of companions on the observed properties of true subgiants. An ISTARY isochrone \citep{istary} for a 9\,Gyr solar metallicity population (dashed red line) clearly demarcates the cool subgiant domain from the field turnoff (orange symbols), and we remove all of these stars from our final sample. However, there is a substantial population of stars close to the main sequence. that appear to be very old subgiants. We can confirm that this is not a truly old population because it is absent far from the main sequence, and seen only in the domain where equal-mass main sequence binaries would appear (below the teal line). This illustrates the frustratingly significant overlap between main sequence photometric binaries and hot subgiants, which we need to identify and remove. Given that it is an area of known equal-mass binary contamination, the stars beneath the teal line are good for validating our tools' effectiveness at identifying these stars. Furthermore, even an MS companion of a true subgiant could impact our mass estimates. We will therefore need to check for companions even in cool subgiants.

Finally, in the bottom panel, we show the domain where the observed abundances must be corrected for dilution by the first dredge-up. Using this, we overlay a heat map showing the detection fraction (\detfrac) as a function of HR diagram position for near-solar-metallicity stars. The domain where dilution exceeds 1.5 dex coincides with the severely depleted region of the superposed detection fraction and is thus a natural region to place the cool boundary of our sample, which we set at 5100 K. 

\subsection{Constraining Random and Systemic Errors in GALAH} \label{1to1}

Reliable lithium measurements depend critically on accurate stellar parameters. Systematic offsets in HR diagram position translate directly into systematic differences in mass due to the tight mass–luminosity relation. We quantify both random parameter uncertainties and systematic model offsets through comparison with peer datasets.

\subsubsection{Random Errors in GALAH Stellar Properties} \label{randomerr}

The median of the errors from the GALAH DR4 parameters in $\feh$, $\teff$, and $\logg$ are $\sigma_{\feh} = 0.059$ dex, $\sigma_{\teff}=69.9$K, and $\sigma_{\logg} = 0.103$, respectively. Additionally, we computed the star-by-star difference between GALAH \citep{galahdr4} and Gaia XPBoost \citep{gaiaxgboost} stellar parameters, and found that the median absolute deviation (MAD) of the Fe/H differences to be $\sigma_{\Delta\feh} = 0.049$ dex, $\teff$ to be $\sigma_{\Delta\teff}=50.2K$, and $\logg$ to be $\sigma_{\Delta\logg} = 0.056$. We thus conclude that the GALAH uncertainties for these quantities are reasonable.

To quantify how these errors factor into our mass interpolation, we propagate the measurement uncertainties in $\loglbol$, $\teff$, $\logg$, $\feh$, and in the bolometric corrections. This yields the expected mass uncertainty due purely to the formal random errors. Using this approach, we find a median YREC mass error of $\sigma_{\rm Med, YREC} = 0.0141$ with a MAD of $0.003$. The GALAH masses yield a similar median error, $\sigma_{\rm Med, GALAH} = 0.0158$, but with a slightly larger scatter, MAD of $0.0035$. These small formal uncertainties reflect the strong main sequence mass-luminosity relationship.

\subsubsection{Systemic Errors in Models: Implications for Mass Estimates}
GALAH masses are independent of our masses, and a comparison of the two is a good measure of systematic uncertainties. As we expect offsets to depend on HR diagram position, we bin our data in slices of $\loglbol$ containing 3000 stars, and further subdivide them at every $ 300 \text{K } \teff$. A small number of outliers can bias the mean statistics, so we used median statistics. Figure \ref{fig:systemic_compare} shows the mass offsets and local mass dispersions. Each row is for data in a 0.1 dex-wide bin in $\feh$, centered at -0.3, 0.0, and +0.3 at the top, middle, and bottom, respectively. The left column shows the median $\mathrm{\Delta M}$ as a function of mass, colored by $\teff$. The middle shows $\mathrm{\Delta M}$ as a function of $\teff$, colored by $\loglbol$. The rightmost column shows the MAD of the mass differences as a function of $\teff$, colored by $\loglbol$. Overall, we see good agreement between the models in hot subgiants. For the large majority of the sample, systematic differences are at or below the 0.05$\msun$ level. There is consistent evidence of a $\teff-$dependent tilt in the mass scale, of order $0.10\msun$ from the cool to hot side, and a comparable stretch in the mass scale from low to high. The largest deviations are seen in the high-mass and cool domain. The random errors in the differences (right columns) are small. We can therefore anticipate that the relative patterns derived for our masses will be topologically similar to what would have been obtained with GALAH ones; our method does allow us to compute dilution factors, which is a major advantage. Systematic errors are larger than random ones. The lithium dip feature could, however, be affected by mass systematics; if the true mass is not captured by the models, one could, in principle, blend contaminants into the sample in a $\teff$-dependent fashion. However, the lithium dip domain appears to be quite flat in $\teff$ ($\mathrm{\Delta M}\leq0.05$), and exhibits a small bias ($\mathrm{\sigma_{\Delta M}\leq0.01}$). We will explore this concern further in Section \ref{lidip_section}. 

In principle, spectroscopic $\logg$ values can provide an independent check on HR diagram-based masses and identify over-luminous binary stars. However, the GALAH $\logg$ values are derived via Bayesian fitting of \textit{Gaia}-based radii to PARSEC isochrones, and thus reflect model-dependent quantities rather than independent spectroscopic measurements — precluding their use as a binary discriminant and necessitating the usage of catalogs. 

\subsection{Sample Decontamination} \label{contaminants}
% \label{GALAH Subgiant Sample Selection}

We apply a sequence of theoretically motivated and dataset-based cuts to yield a clean single-star subgiant sample. 
\begin{figure}
    \centering
    \includegraphics[width=1\linewidth]{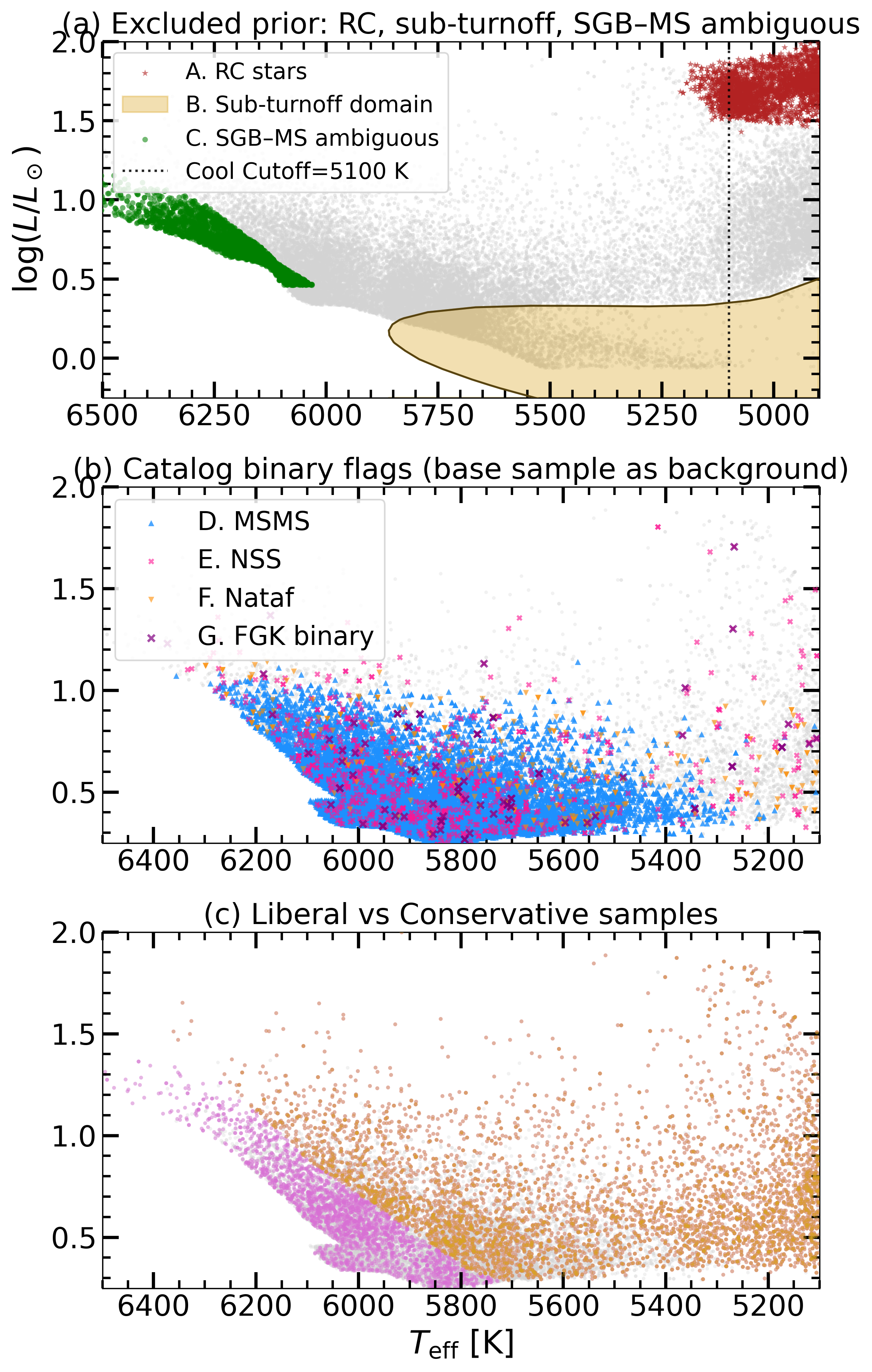}
    \caption{GALAH subgiants for $\feh = 0$ illustrating the different stages of sample decontamination. \textbf{TOP}: The initial removal from Figure \ref{fig:catalog_sample_compare}, showing the domains we initially exclude (including the $\teff=5100$K cut) \textbf{MIDDLE}: Stars removed by binary datasets. Group D removes the most, with useful contributions from Groups E-G. \textbf{BOTTOM}: We present our sample with and without the Equal-Mass Binary Domain (the region $\log(2)$ dex in $\loglbol$ from the main sequence).
    }
    \label{fig:phasespace_binary}
\end{figure}
The top panel of Figure \ref{fig:phasespace_binary} illustrates domains that we exclude completely from our sample: the area of SGB-MS phase ambiguity, stars with $\teff\leq 5100$K, the subturnoff, and the region that overlaps with the red clump. The middle panel demonstrates the population of literature binaries that we remove. The bottom panel shows our remaining sample in two domains. The orange stars are the main sample. The pink ones are in the domain where unrecognized main sequence binaries could contaminate the sample.
\begin{table}[ht]
\centering
\footnotesize
\caption{Summary of the sample construction \& cuts.}
\label{tab:flag_removal_summary}
\begin{tabular}{@{}p{0.5\linewidth}r@{}}
\toprule
\multicolumn{2}{c}{\textbf{Stage 1 --- Quality cuts}} \\
\midrule
\multicolumn{2}{@{}p{\linewidth}@{}}{\textit{Cuts applied: (A) RC stars; (B) sub-turnoff; (C) SGB--MS ambiguous; $T_{\rm eff} < 5100$\,K}} \\
\midrule
Input & 165{,}660 \\
Output & 94{,}899\ (57.29\%) \\
\toprule
\multicolumn{2}{c}{\textbf{Stage 2 --- Catalog binary removal}} \\
\midrule
\multicolumn{2}{@{}p{\linewidth}@{}}{\textit{Cuts applied: (D) MSMS catalog binary; (E) Gaia NSS binary; (F) Nataf et al.\ binary; (G) GALAH FGK binary catalog; (H) Equal-mass binary domain}} \\
\midrule
Input & 94{,}899 \\
Liberal output (excl.\ H) & 39{,}417\ (41.54\%) \\
Conservative output (incl.\ H) & 28{,}434\ (29.96\%) \\
\toprule
\multicolumn{2}{c}{\textbf{Stage 3 --- Hot-side cut}} \\
\midrule
\multicolumn{2}{@{}p{\linewidth}@{}}{\textit{Cuts applied (cool plateau only): $T_{\rm eff} \leq 5800$\,K}} \\
\midrule
Liberal input & 39{,}417 \\
Liberal output (excl.\ H) & 28{,}434\ (29.96\% of initial) \\
Conservative input & 28{,}434 \\
Conservative output (incl.\ H) & 22{,}442\ (23.65\% of initial) \\
\bottomrule
\end{tabular}
\end{table}

Each binary catalog utilizes different criteria for identifying and removing binaries. The Main Sequence Million Star (MSMS) Binary catalog \citep{msmsbinarycat} of 35,000,000 stars (and $\sim14,000,000$ equal-mass binary candidates) within 1 kpc, uses Gaia XP spectra \citep{gaiaxgboost}, a neural net trained on SED spectra and CMD positions of single stars, and PARSEC isochrones \citep{parsec1} to specifically identify the equal-mass binary stars that have scattered into the subgiant branch. Their statistical weighting of XP spectra, parallax-informed distance, and CMD placement relies purely on observables. We cross-match our GALAH sample with stars flagged as high-probability binaries ($\text{Predicted Class} = 1$), defining our \emph{\textbf{Group\,D}} stars in Figure~\ref{fig:phasespace_binary}. Binaries can also be identified via precise photometric, radial velocity, and kinematic measurements. We remove stars cross-matched in the Gaia DR3 Non-Single Stars (NSS) catalog \citep{gaiadr3}, and classify these as our \emph{\textbf{Group E}} stars. Using the method of \cite{chancemethod}, \cite{natafcatalog} identifies binaries by statistical correlation between Gaia DR3 RV errors and the broadening produced by Spectroscopic binaries. These comprise our \emph{\textbf{Group F}} stars in Figure \ref{fig:phasespace_binary}. \citet{natafcatalog} is sensitive to cooler subgiants, thereby complementing the \citet{msmsbinarycat} catalog optimized for the turnoff domain.  \cite{lyrabinaries} showed that RV variability and a poor astrometric solution can also be useful binary diagnostics. We thus remove stars that are cross-matched by ID in the Gaia RV variable and the GALAH FGK binaries datasets \citep{gaiavari_all, galahfgkbinary} (\emph{\textbf{Group\,G}}).

\begin{figure}
\centering
    \includegraphics[width=1.0\linewidth]{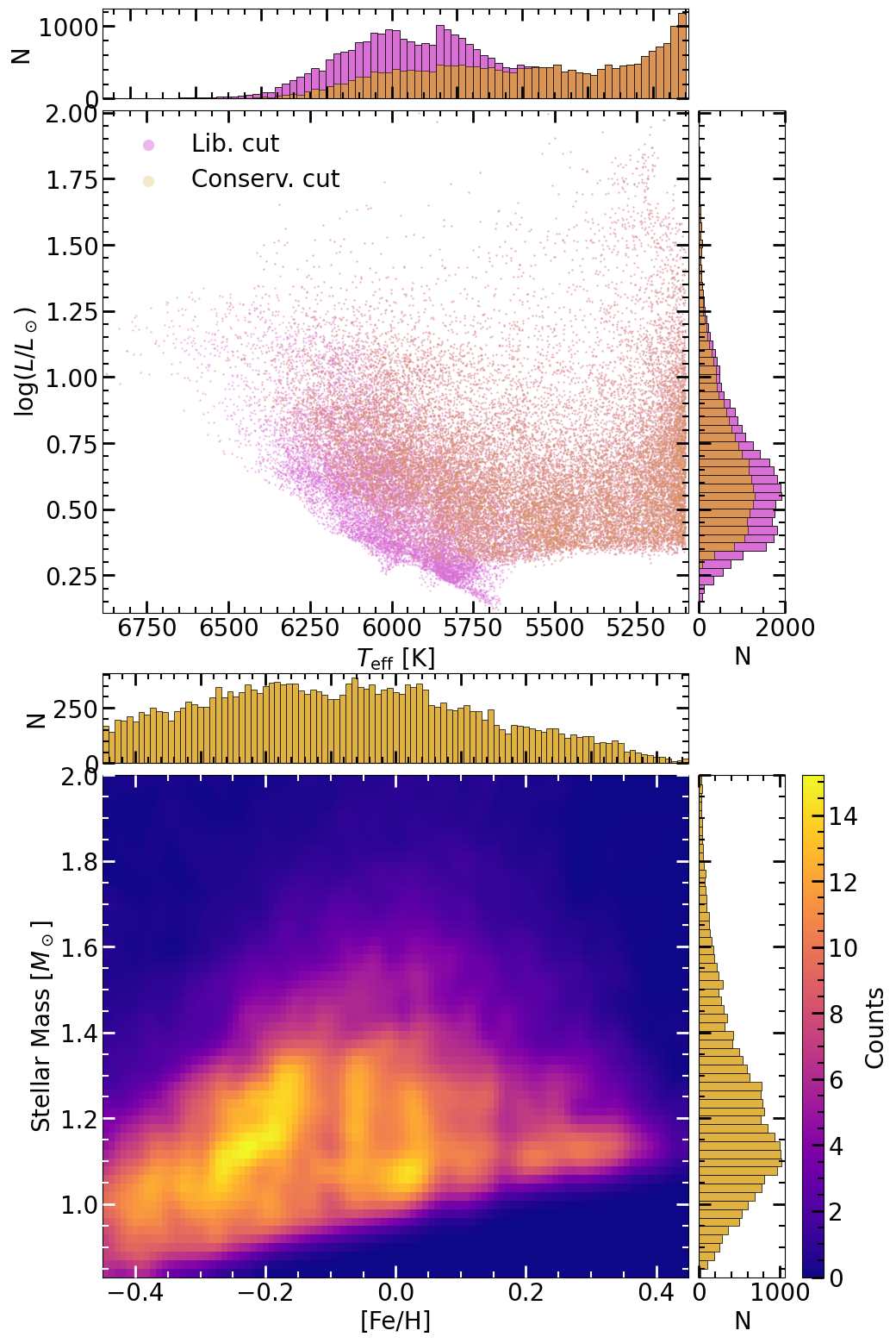}
    \caption{GALAH subgiant
    sample. All panels exclude red clump stars, sub-turnoff
    binaries, SGB--MS ambiguous stars, and stars with $T_{\mathrm{eff}} < 5100$~K, giving initial sample $N = N_{\mathrm{original}}$ (annotated, right panel). \textbf{TOP:} HR diagram for the \textit{liberal cut} (orchid; initial sample
    minus all catalog binary flags; $N = N_{\mathrm{liberal}}$) and the
    \textit{conservative cut} (goldenrod; liberal cut minus
    \textbf{\textit{Group H}}; $N = N_{\mathrm{conservative}}$). \textbf{BOTTOM:} $[\mathrm{Fe/H}]$--mass density for the conservative cut.
    Histograms accompany both panels.}
    \label{fig:feh_mass_comparison_full_vs_clean}
\end{figure}
\begin{figure*}
    \centering
    \includegraphics[width=1\linewidth]{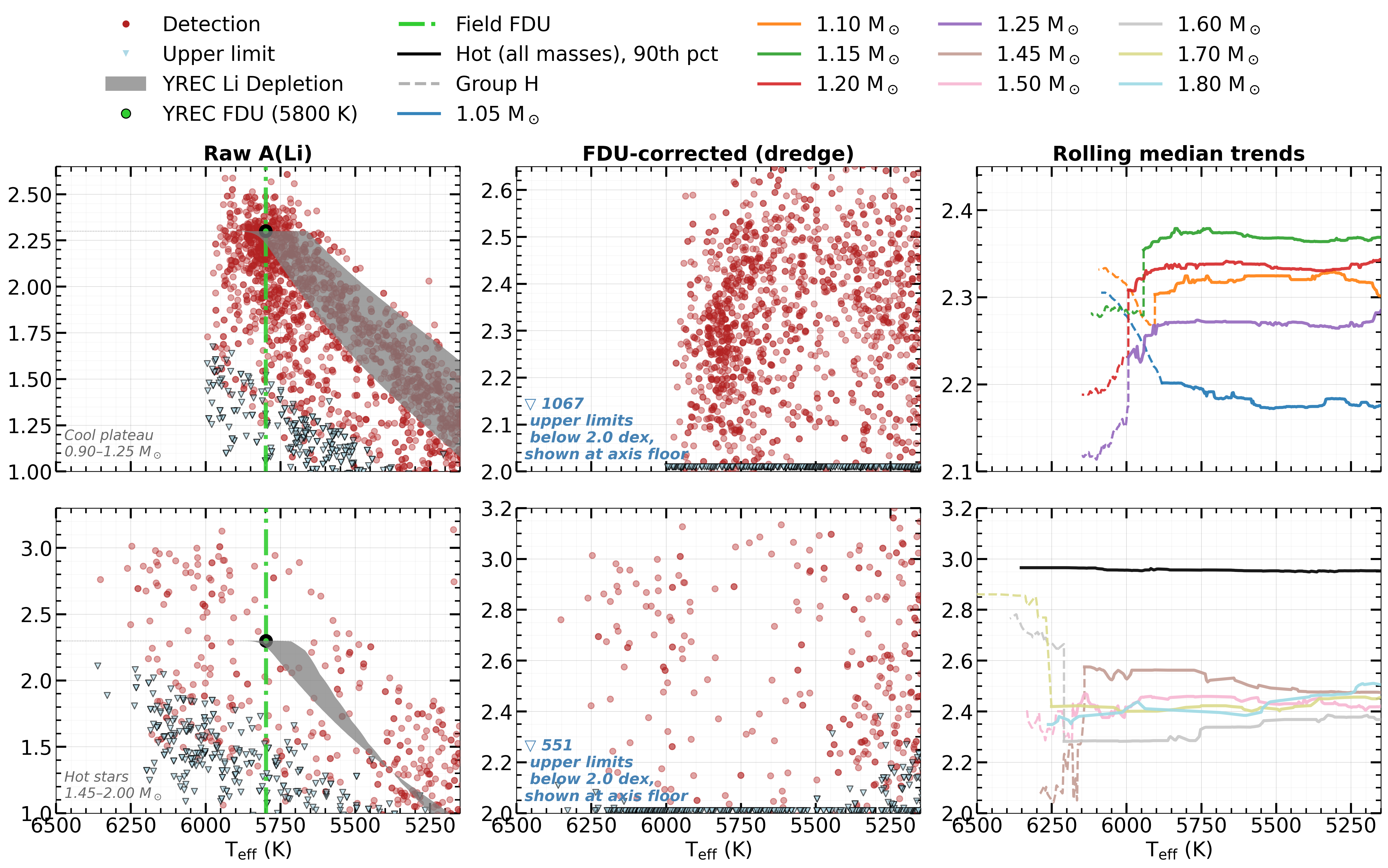}
    \caption{$\teff$ vs. $\ali$ abundance for the solar-metallicity ($\feh_{\odot} \pm0.05$) subgiants from our conservative cut in two mass regimes: below the lithium dip (\textbf{top group}), and above the lithium dip (\textbf{bottom group}). Within each group, the \textbf{leftmost panel} shows raw $\ali$ detections (red circles) and upper limits (blue triangles), with the gray shaded regions showing the range of lithium depletions for the mass range as a function of temperature. The starting point is defined at the median lithium in each mass range (2.3 for low and high, respectively). The lime-green circle marks the canonical FDU onset at $\teff = 5800$\, K. The \textbf{panel second from the left} shows the dredge-up-corrected $\ali$ after applying individual YREC dilution factors. The \textbf{third panel shows} the rolling 50th percentile in 100 star bins for detections and upper limits at different masses. The smaller dashed lines extending from them represent the rolling trends extended into the excluded Group\,H domain. The solid black line represents the 90th percentile/upper envelope of the whole massive star sample.}
    \label{fig:solarfeh_dredgeliteff}
\end{figure*}
In addition to our catalog binaries, we define (\emph{\textbf{Group\,H}}) as $0.3$ dex in $\loglbol$ above the main sequence as the domain where equal-mass binaries overlap with the subgiant branch. We use the subturnoff within \textbf{\textit{Group\,H}} as a control to validate our binary recovery, as theoretical models predict that this region is comprised solely of contaminants. \textit{Inside} the \textbf{\textit{Group\,H subturnoff}}, we can match up to $\approx 68\%$ of stars to \textbf{\textit{Groups\,B-G}}, indicating a $\sim32\%$ contamination fraction not captured by literature catalogs. Looking generally at the \textbf{\textit{Group\,H}} domain \textit{as a whole}, we have already excluded $\approx72\%$ of the total GALAH sample from \textbf{\textit{Groups\,B-G}}. This very high rate indicates that binary contamination in this domain is quite severe, as expected from stellar population models. We therefore divide our sample into a  \textbf{\textit{liberal}} (including the \textbf{\textit{Group\,H}}) and \textbf{\textit{conservative}} (excluding \textbf{\textit{Group\,H}}) \textbf{\textit{samples}}. For our analysis, we use the \textbf{\textit{conservative}} sample to maximize confidence that our sample contains negligible contaminants (and make both cuts public). 

We tabulate the results of our cuts in Table \ref{tab:flag_removal_summary}. The resulting final sample, shown in Figure \ref{fig:feh_mass_comparison_full_vs_clean}, has a broad metallicity distribution, which tails off at the high $\feh$ range. The mass distribution is strongly weighted toward low mass, as expected from lifetime and IMF arguments \citep{kroupaimf}. There is a sharp lower edge in mass, which is metallicity-dependent; this represents our 9\,Gyr age cut for the sample \citep{massmetfunction, stellarmultiplicity, xiangrixthickdisk}. Stars from the thick disk are outside our scope, but could fall within our sample. While most lie outside the metallicity domain of our sample, our age cut affirms we've excluded the rest. With binaries and contaminants constrained, we can now correct for the effects of the dredge-up.

\subsection{Applying the First-Dredge Up Correction} \label{dredgeupsection}
Using \texttt{Kiauhoku}, we interpolate the YREC grid to compute the absolute lithium depletion along the track matching the star's mass and 
$\feh$, from the turnoff to the star's $\teff$. We then apply this star-by-star dredge-up correction factor to the current abundance. This is defined as the difference between the lithium abundance at the turnoff in Figure \ref{fig:catalog_sample_compare} for a model of that mass and metallicity and the abundance at the star's current HR Diagram position. We now examine other physical effects that may affect the turnoff abundance trends.

\subsection{Crafting the Turnoff Abundance: Constraining the Sample of Dredge-Up Corrected
Stars}\label{single_binary}

It is well known that lithium is diluted as stars develop deep surface convection zones during the first dredge-up. However, if the temperature at the base of the convection zone becomes high enough, lithium can also be burned and directly destroyed. The latter phenomenon becomes important in cool subgiants with low mass and high metallicity \citep{iben1967fdupostms_b, lambertfdu,kailimarc2025}. We use the comparison of data trends with theoretical predictions to set physically motivated hot and cool bounds for our sample and explore whether standard models correctly predict the trends seen in the data.

Mixing processes can reduce the abundances faster than predicted by standard models because mixing changes the internal abundance profile. Gravitational settling \citep{sestrand2005, Soutom67} or radiative acceleration \citep{Michaud1970, abt1979, Abt1995,vauclair2005, Abt_2017} can also modify the internal abundance profile, which can change the predicted surface abundance trends. For gravitational settling, this can manifest as a recovery in lithium just past the main sequence, as lithium can sink just below the surface but not burn \citep{Soutom67}. For radiative acceleration, lithium is concentrated and only preserved in a thin surface layer. This results in more severe lithium dilution for cooler stars than standard models would predict \citep{Michaud1970,michauddiffusionfstars,abt1979, Abt_2017, Abt1995}.
\begin{figure*}
    \centering
    \includegraphics[width=1\linewidth]{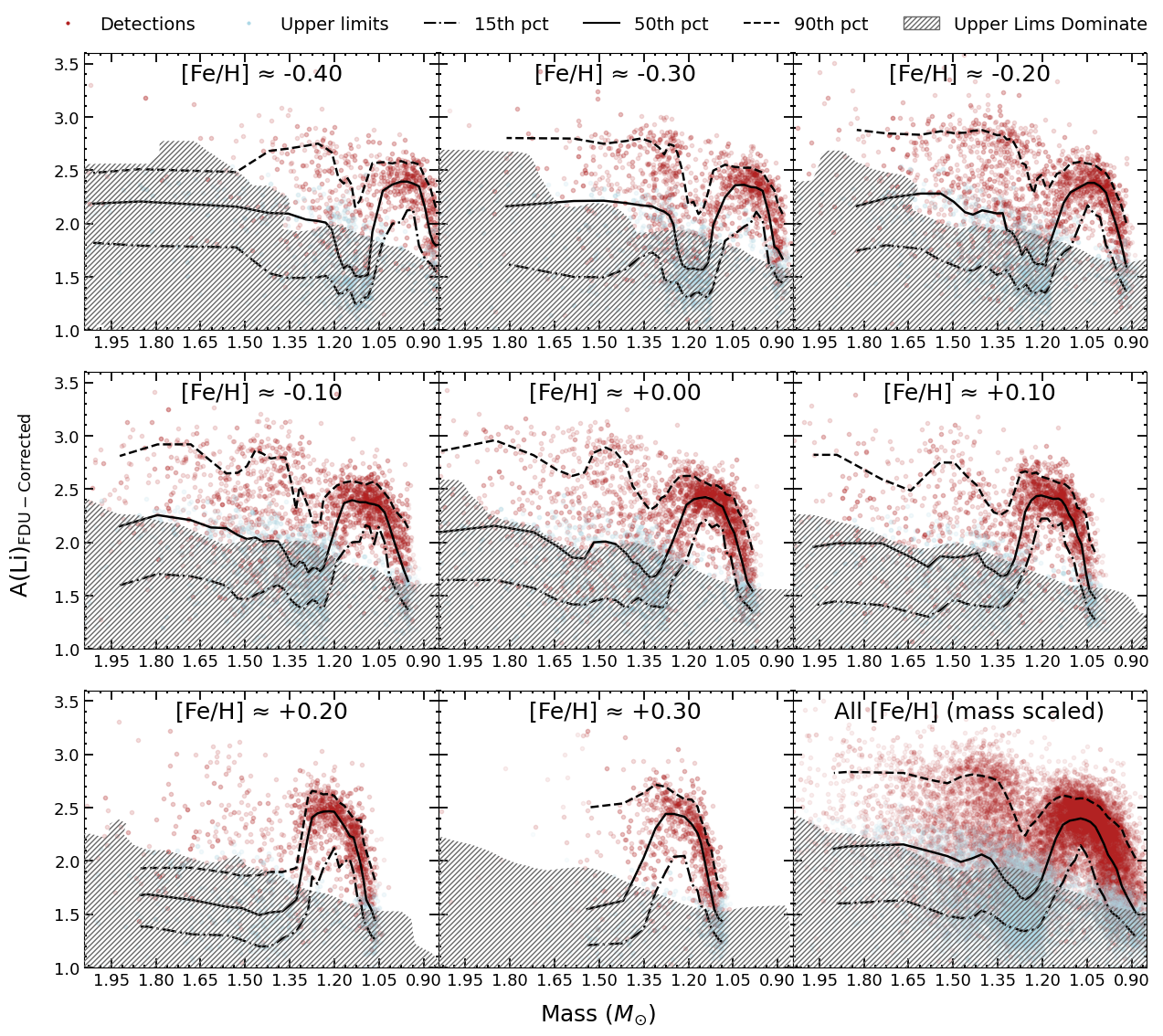}
    \caption{The inferred MS turnoff lithium abundances for our conservative sample. Detections (red) and upper limits (blue) are shown as a function of mass and metallicity. Each sub-panel depicts a bin with a different central $\feh\pm0.05$. Dashed and solid lines represent rank-order percentiles of the lithium abundance using 200 star bins. The bottom right shows the entire sample, with a metallicity-dependent mass shift that aligns the lithium dip (see the text for details). The shaded crosshatch represents the region where $\geq50\%$ of stars are upper limits.
    }
    \label{fig:massdredge}
\end{figure*}
\begin{figure*}
    \centering
    \includegraphics[width=1\linewidth]{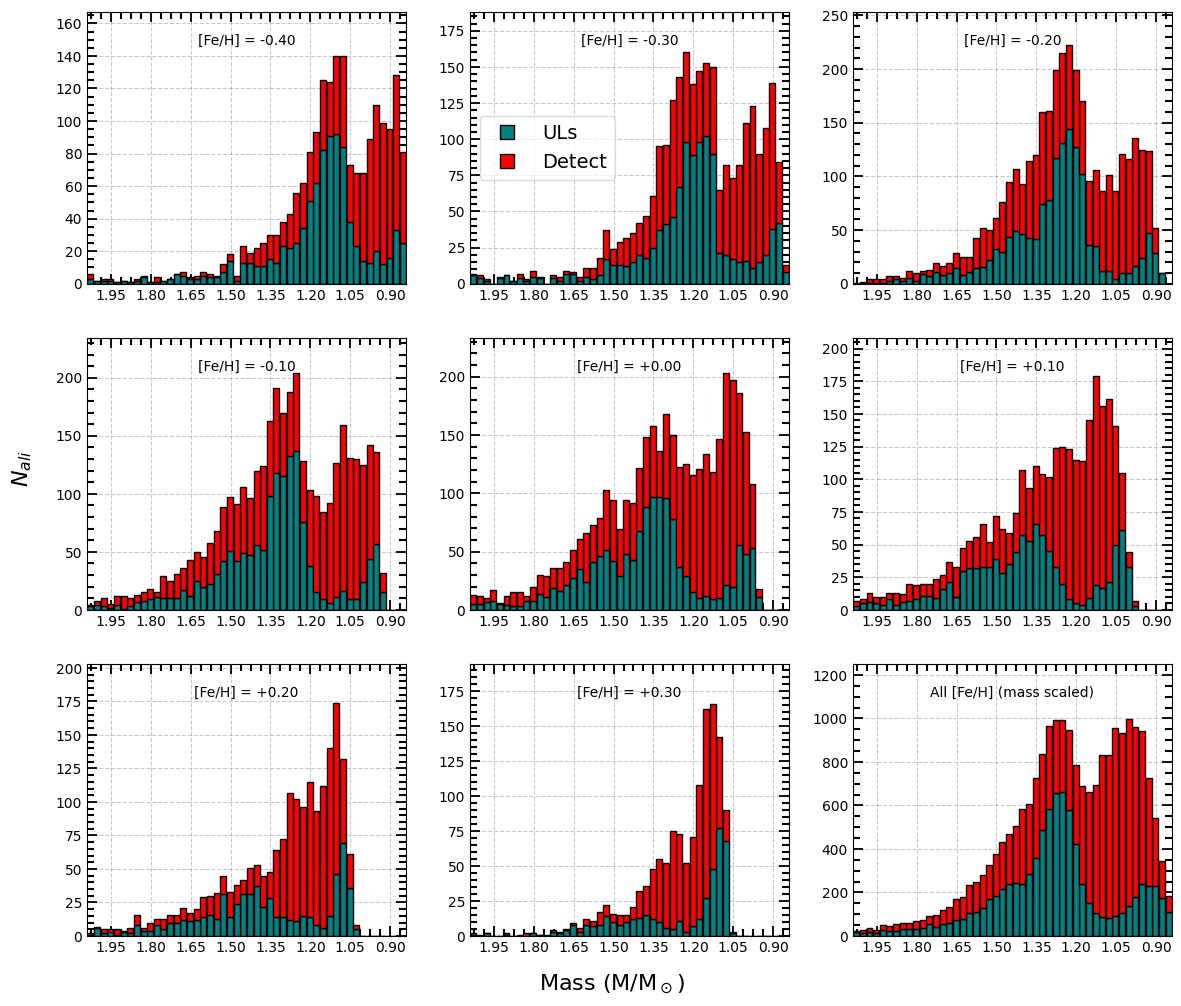}
    \caption{ Detections (red) are stacked on the upper limits green) By mass to denote where either regime dominates in mass space. The proportion of detections is dominant in the cool plateau, shifting to an upper-limit dominance through the lithium dip and into the hot side of the dip. The bottom right denotes the whole sample, empirically scaled by $\feh$ to the same quantity $\mscaled$ (which we elaborate in the following section).}
    \label{fig:3x3stackedhistplot}
\end{figure*}

With these mixing and diffusion processes in mind, we compare theory and data for the $\feh$ low-mass ($1.05\leq \msun \leq 1.20$, top panels) and high-mass ($1.45\leq\msun\leq1.80$, bottom panels), in Figure \ref{fig:solarfeh_dredgeliteff}. We do not show the lithium dip because the sample is largely composed of upper limits, where we cannot test directly for consistency with first dredge-up predictions. The left column shows the raw data and the range of theoretically predicted depletion factors for the mass ranges in question. The middle column illustrates a zoomed-in representation of detrended data, focused on the median behavior. Rolling median trends at fixed mass are shown in the right column. The overall trends here elegantly mirror theoretical predictions. 

For the cool plateau stars, the median lithium abundance at fixed mass does not deviate by more than $\sim0.02$ dex from $5100\text{K} \leq \teff \leq 5800\,\text{K}$. However, we do see clear trends emerging hotter than 5800\,K, even for our conservative sample. These are statistically significant ($\Delta\ali\sim0.046\pm0.015$) and seen in all bins. The sense is that the apparent abundances are lower for hotter stars, which is predicted by models that include gravitational settling and thermal diffusion in \cite{chaboyerrimm67gkstars,Soutom67}. Although we refer to these as MS turnoff, they may therefore more properly be thought of as the abundances just before the onset of dredge-up. 

The Group\,H (equal-mass binary) domain, shown by mass bin with dashed lines, shows stronger trends. We observe a sharp $\sim0.2$ dex change relative to the median of the cool stars. The likely culprits are plateau binaries scattering into the depleted region near the lithium dip, or already-depleted stars scattering upward into the plateau. These trends bolster the idea that binary contamination is still relevant in this domain, even after the literature binaries are removed. 

For the massive stars, equal-mass binaries can enter the sample at a higher $\teff$ than they do for cool stars as well, becoming a factor only above 6250\,K. There are hints of trends above 5800\,K for the bins below $\leq1.7\msun$ that are consistent with the low-mass pattern. Some care is required here because of the large upper-limit fraction; the median hot-star measurement straddles detections and upper limits. Cool upper limits are more stringent than hot ones, which can inject artificial trends that lack statistical meaning. The top 10$\%$ of stars (indicated by the solid black line) are much flatter in trend, indicating that our technique is not imposing large $\teff$-dependent systematics. We also observe significant $\teff$ trends in the Group\,H domain, due to similar phenomena to the cool plateau case. 

Informed by these trends, we make our final cut by excluding the Group\,H domain from our analysis. Our finished sample only includes stars between $5100 \leq \teff \leq 5800$\,K below the dip, and extends this to $5100 \leq \teff \leq 6250$\,K above the dip.

\section{The Turnoff Lithium Abundance Pattern} \label{lithium pattern}

We are now in a position to assess the lithium pattern as a function of mass and metallicity. In Figure \ref{fig:massdredge}, we divide our data into 0.1\,dex metallicity bins and plot lithium as a function of mass. Upper limits are in blue, and detections are in red. To characterize the trends, we rank-order the data in mass and plot the 15th, 50th, and 90th percentiles of a rolling 200-star sample for detections and upper limits. The domain where upper limits account for $\geq 50\%$ of abundances is crosshatched grey. We begin with the solar metallicity panels. Solar-mass stars leave the main sequence with a range in lithium, typically higher than the solar value of $\ali_{\odot}\approx1.0$. The median abundances rise to a cool plateau around 1.1$\msun$, with a high detection fraction. We will discuss this in Section \ref{coolplat_section}. At higher mass, there is a sudden drop to a population dominated by upper limits across a narrow mass range ($\sim 0.1 \msun$ wide); Figure \ref{fig:massdredge} shows that this drop is at least an order of magnitude, and Figure \ref{fig:3x3stackedhistplot} shows the striking rise in the upper limit density in this mass range. The c-shaped cool side and lithium dip mirror the open cluster distribution \citep{Balachanrdanm67, paceclustm67, pleiadesbouvier, Soutom67}. Upper limits dominate approaching the Kraft break and $\lesssim1.0\msun$.
Beyond the cool plateau, however, the pattern exhibits striking differences from the cluster pattern. There is a large subset of detections in the lithium dip, which we will return to in Section \ref{lidip_section}.  The upper envelope of the high-mass star distribution rises sharply, similar to what is seen in clusters. It approaches, but does not reach the solar meteoric value of $\ali_\text{meteoritic}=3.3$. However, the high-mass distribution is extremely broad; the median $\ali$ straddles the detection-upper limit boundary with an intrinsic width of at least $1.5$\,dex across a wide range of masses. Although some scatter is evident for higher-mass cluster stars as well, the distribution there is tighter and systematically higher than the field data set. We discuss this population in Section \ref{hotforksection}.

Turning to the $\feh$ trends, the cool plateau depends only weakly on metallicity, but the mean mass clearly shifts to higher mass at higher metallicity (Figures \ref{fig:massdredge}-\ref{fig:3x3stackedhistplot}). Across metallicities, the topology of the lithium dip is similar, including width, with a mass shift that parallels that of the cool plateau. The high-mass pattern is self-similar from -0.3 to +0.1; at higher and lower metallicities the data become dominated by upper limits, and the sample size shrinks significantly. We will explore these interesting trends in Section \ref{hotforksection}. Thus, the lithium pattern hotter than the dip is akin to a ``lithium fork": the upper envelope of lithium rises dramatically hotter than the dip, but this coincides with a large, highly depleted cohort. It is apparent, then, that the field stars may present more features to the lithium dip than constraints on cluster observations allow.

The strong mass-dependent trends clearly indicate that different physical processes dominate in stars below, in, and above the lithium dip. We therefore elect to take our dredge-up corrected sample and divide it using the lithium dip as the boundary, and analyze the structure of the pattern piecewise.

\begin{figure*}
    \centering
    \includegraphics[width=.85\linewidth]{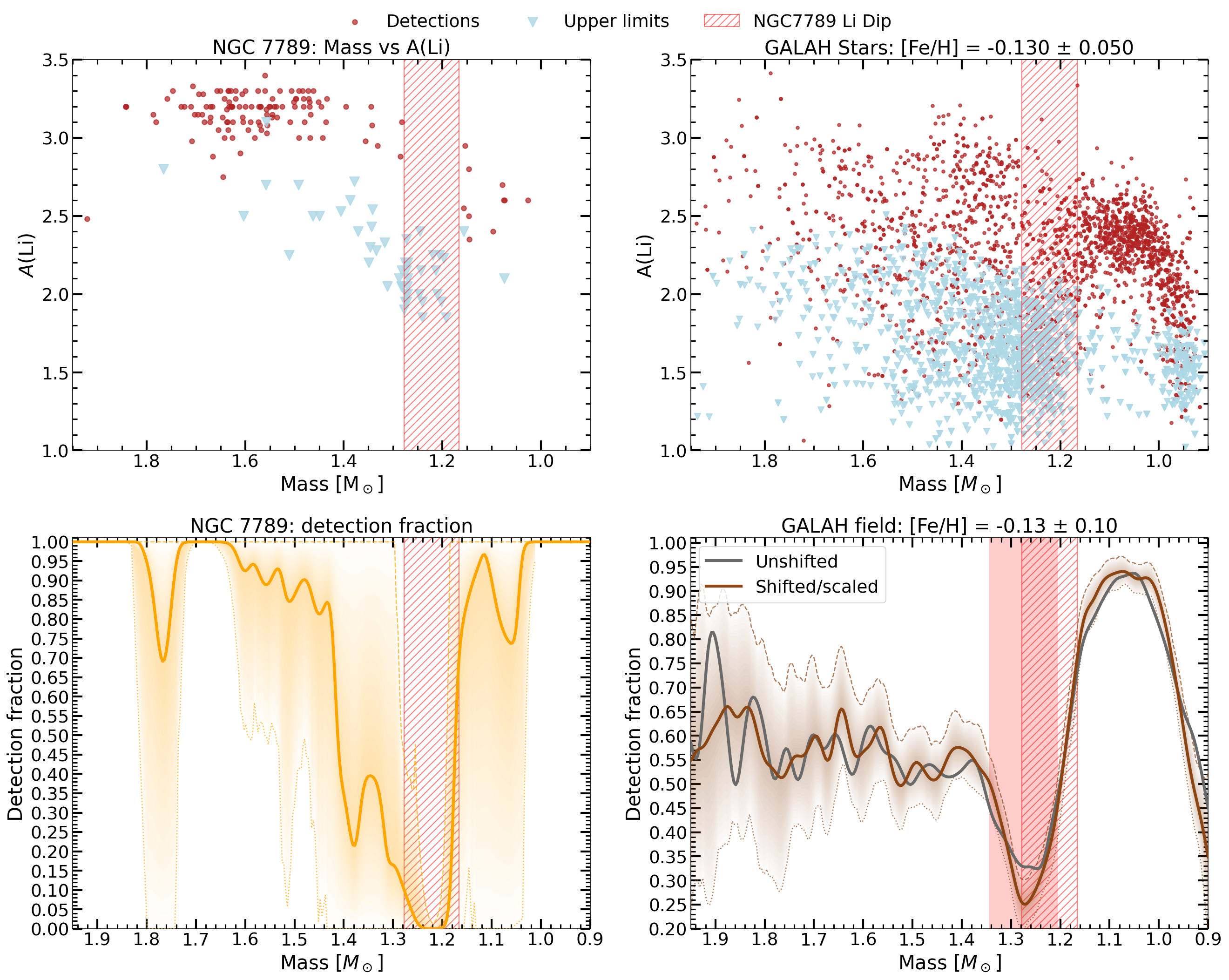}
        \caption{Comparison of the lithium dip in NGC~7789 and GALAH field stars at matching metallicity.
\textbf{Top Left:} Mass vs $A(\mathrm{lithium})$ for NGC~7789 from \cite{ngc7789lidip}, with shading
marking the mass range of the lithium dip.
\textbf{Top Right:} GALAH subgiant stars at the metallicity of NGC~7789, with the lithium dip
mass range shaded. \textbf{Bottom Left:} 
The KDE of the Detection fraction as a function of mass. The gold solid line represents the detection fraction of the cluster, with the lithium dip shaded in the pink-red crosshatch. The gold shading of the detection fraction curve represents the maximum and minimum possible smearing of the lithium dip feature by mass errors, which we approximate using $\sqrt{N_{stars}}$ in each bin. 
\textbf{Bottom Right:} The same axes as the left, except we add the additional curve showing the effect of removing $\feh$ binning effects. The blue curve represents the detection fraction KDE using native $M_{\text{YREC}}$ masses, and the brown curve shows it by empirically scaling the masses of all stars in the 0.1 dex-wide bin to the central $\feh_{NGC7789}$. We also see the shaded brown band, representing the maximum and minimum possible smearing of the lithium dip feature. Here we perform 1000 random drawings of 2000 stars and compute the detection fraction as a function of mass. Then we take the $\pm5\sigma$ of the resulting distributions. We show the NGC 7789 lithium dip in crosshatch and the field dip shaded in pink.}
    \label{fig:cluster_field_dip}
\end{figure*}
\subsection{The Lithium Dip: Calculation, Bounds, \& Properties} \label{lidip_section}
The lithium dip \citep{bosegaard1986,bosegaard1987} poses a dramatic, clear challenge for standard models. In contradiction to expectations, the lithium abundances in the mid-F region of the main sequence plunge by more than a factor of 30 over a narrow temperature range in under a Gyr \citep{bosegaard1986,bosegaard1987,ngc188rotmixlidip, 35myrstellarsnake}. On the main sequence, lithium is undetected in the dip with stringent upper limits. Furthermore, their post-main sequence descendants in clusters also exhibit this, indicating that lithium is destroyed rather than hidden in the dip stars \citep{bosegaard1986, bosegaard1987, stillsdeliyannis2000, ngc188rotmixlidip, ngc7789lidip}. In the GALAH field sample, \citet{gao2020} found a high density of strict upper limits within a narrow mass range. In turn, we see the inherited lithium dip in our subgiant sample (Figures~\ref{fig:massdredge} and \ref{fig:3x3stackedhistplot}), prompting our own calculation of the width, bounds, and central mass of the field lithium dip.

Star cluster data is a natural reference point for locating the lithium dip. We use ($\mathrm{M_{v}<14}$) data for NGC\,7789, shown in the top left panel of Figure \ref{fig:cluster_field_dip}. The crosshatched region is the lithium dip defined in \citep{ngc7789lidip}, $15.2\leq \mathrm{M_{v}} \leq 16.5$. We convert to mass using their given $\log(\mathrm{g})$ and our YREC tracks. The top right panel shows the field subgiants at a similar metallicity ($-0.18 < \feh < -0.08$), with the NGC 7789 lithium dip mass range overlaid for reference again. Although the field sample here is clearly dominated by upper limits, the GALAH sample has a sizable sub-population of detections within the lithium dip region. There is also a striking contrast between cluster and field in the higher mass sample, which we will explore in Section \ref{hotforksection}. The differences in the dip morphology could have several causes. It might be a real population difference. However, it could instead reflect a contamination of the sample by outliers, such as unrecognized binaries. If the dip feature is narrow enough, it might be blurred by binning effects, systematics, or random errors in mass and metallicity. To distinguish between these possibilities, we therefore more precisely define the location, width, and metallicity dependence of the lithium dip itself. 

We use a detection fraction, the ratio of detections to the sum of our detections and upper limits, as a function of mass for our figure of merit for defining the lithium dip. We define the center as the minimum in detection fraction, and the width is defined as the mass domain where the detection fraction drops below 0.4. An important consideration for doing this is that, as seen in Figure \ref{fig:3x3stackedhistplot}, the location of the lithium dip is clearly sensitive to metallicity. Even within a 0.1\,dex metallicity range, we can therefore anticipate that the location and depth of the dip will be blurred by binning effects. Random mass errors may also scatter stars on the dip edges into the dip itself. We begin by computing the detection fraction as a function of mass by creating a kernel density estimation\,(KDE), with a mass width equal to the median mass error $\sigma_{M_{YREC}} = 0.0141$. We have verified that smaller mass bins do not change the general findings.

We perform this exercise for a range of metallicity bins ($-0.4\leq\feh\leq+0.4$), centered on every 0.1\,dex with $\pm0.05$ bins. We identify a lithium dip at all metallicities, the characteristic mass of which also increases with $\feh$. Motivated by the resulting trends across $\feh$, summarized in Table \ref{tab:lidip}, we construct a scaled mass proxy for each star:
\begin{equation}
\label{equ:massscald}
M_\mathrm{scaled} = M_* + (k=0.505) \left(\mathrm{[Fe/H]}_\mathrm{NGC\,7789} - \mathrm{[Fe/H]}_*\right)
\end{equation}
Our goal is twofold. We want a prescription that characterizes the broad metallicity trends. But we also want a mechanism to correct for the known blurring effect of the metallicity range within a bin. We vary the constant $k$ in Equation \ref{equ:massscald} over the range $0.0 \leq k \leq 10$. Then we adopt the value that minimizes the detection fraction within the mass range of the lithium dip -- resulting in a minimum detection fraction of $\sim25\%$. The results for NGC\,7789 and the GALAH field stars are shown in the bottom left and right panels, respectively, of Figure \ref{fig:cluster_field_dip}. 

\begin{figure}
    \centering
    \includegraphics[width=1\linewidth]{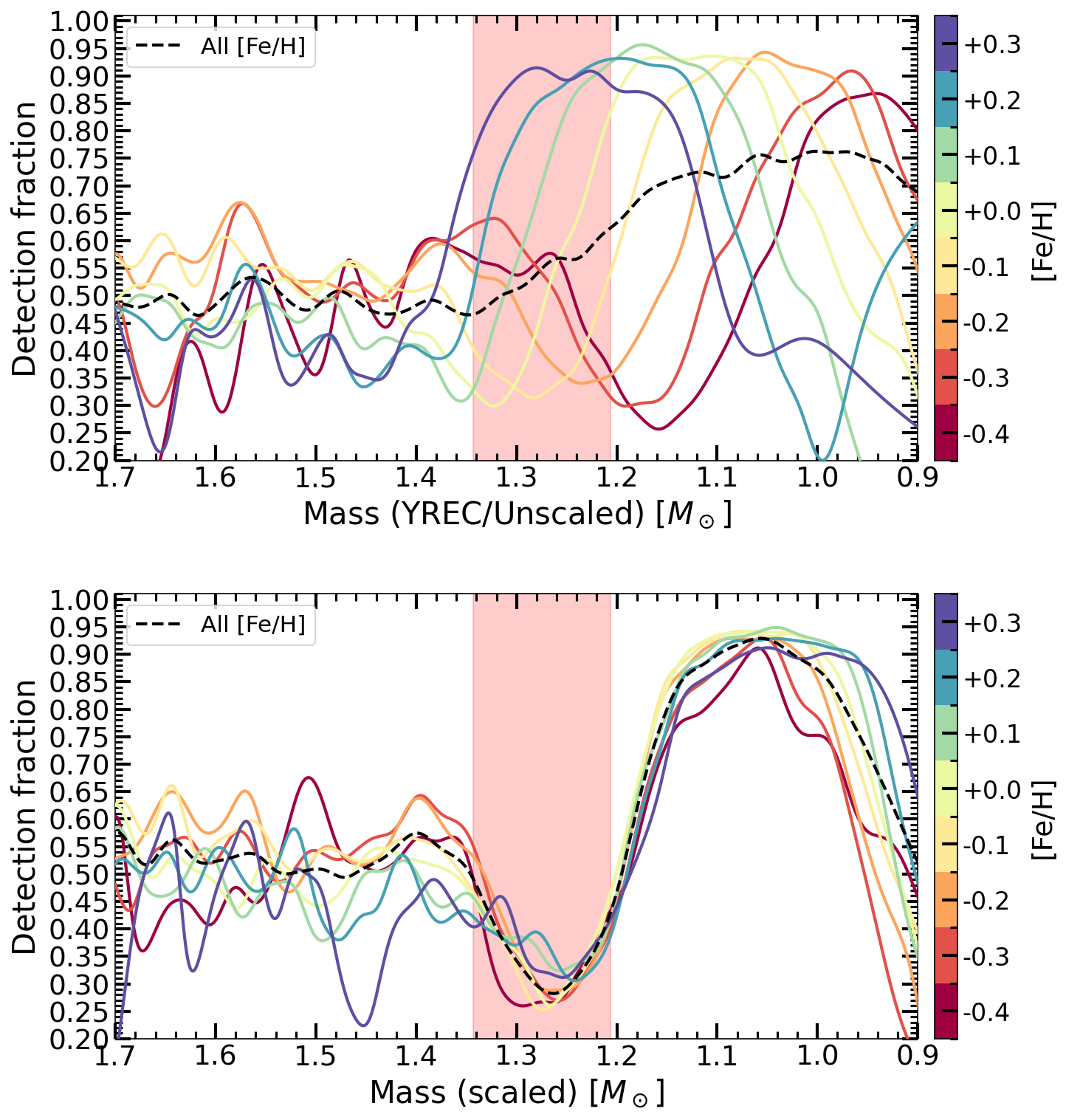}
    \caption{\textbf{TOP:} Detection fraction as a function of YREC-computed mass, ranging from $-0.4\leq\feh\leq+0.4$, with $\feh\pm0.05$ on either side of the bins. \textbf{BOTTOM:} Detection fraction as a function of $\mscaled$, with a dashed black line showing the rolling detection fraction for the whole sample. In both panels, the shaded red mass range represents the lithium dip range computed by Equation \ref{equ:massscald}.}
    \label{fig:detec_frac_dip}
\end{figure}

\begin{table}[h]
\centering
\caption{Lithium dip properties as a function of metallicity, computed from a KDE detection fraction as a function of mass. These correspond well with the identified trends in Figures \ref{fig:massdredge}-\ref{fig:3x3stackedhistplot}.}
\label{tab:lidip}
\begin{tabular}{l  c  c  c  c  c  c}
\hline
{\feh} && Low bound && Central Mass && High bound \\
(dex)  && ($M_\odot$) && ($M_\odot$) && ($M_\odot$) \\
\hline
$-0.40$ && $1.07\pm0.016$ && $1.13\pm0.017$ && $1.21\pm0.019$ \\
$-0.30$ && $1.12\pm0.018$ && $1.18\pm0.017$ && $1.26\pm0.020$ \\
$-0.20$ && $1.17\pm0.021$ && $1.23\pm0.018$ && $1.31\pm0.022$ \\
$-0.10$ && $1.22\pm0.022$ && $1.28\pm0.022$ && $1.36\pm0.026$ \\
$+0.00$ && $1.27\pm0.023$ && $1.33\pm0.024$ && $1.41\pm0.028$ \\
$+0.10$ && $1.32\pm0.022$ && $1.38\pm0.023$ && $1.46\pm0.027$ \\
$+0.20$ && $1.37\pm0.022$ && $1.43\pm0.026$ && $1.51\pm0.027$ \\
$+0.30$ && $1.42\pm0.021$ && $1.48\pm0.027$ && $1.56\pm0.025$ \\
$+0.40$ && $1.47\pm0.021$ && $1.53\pm0.025$ && $1.61\pm0.030$ \\
\hline
\end{tabular}
\end{table}
From this method, we can quantify the effects of metallicity binning and random mass scatter. Binning effects are clearly visible. For the GALAH cluster analogs and other $\feh$, removing these effects deepens the feature by $\sim0.1$. Figure \ref{fig:detec_frac_dip} corroborates this result for the whole sample. The black dashed median line is clearly smeared by these effects in the top panel and resolves the clear dip, plateau, and rise in the bottom. The shift is also visible in the detection fraction curves in the top panel, which corresponds to Table \ref{tab:lidip}. Therefore, while binning clearly impacts the depth of the lithium dip feature, even careful attempts to minimize this do not remove it. 

\begin{figure}
    \centering
    \includegraphics[width=1\linewidth]{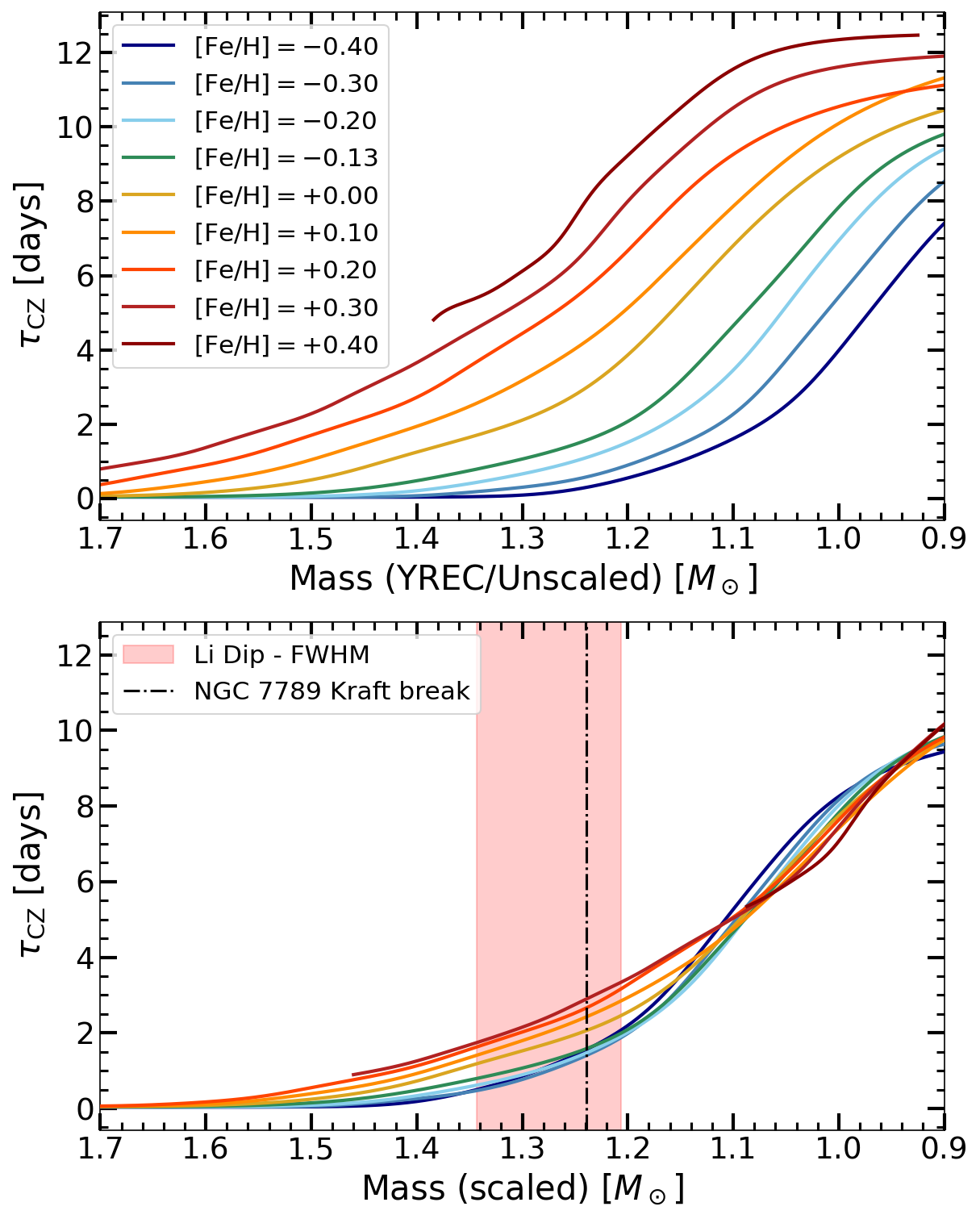}
    \caption{\textbf{TOP:} $\tau_{CZ}$ (days) vs.\ YREC mass; each colored line is the median theoretical $\tau_{CZ}$ for field stars at a given $\feh$, using their YREC-predicted mass. \textbf{BOTTOM:} The same as the above but as a function of $\mscaled$, showing that when shifted, the change between slow and rapid $\tau_{CZ}$ coincides with the lithium dip.}
    \label{fig:taucztest}
\end{figure}

So far, we have defined the lithium dip in purely empirical terms. Physically, it coincides with the transition from radiative to convective envelopes and the accompanying Kraft break — the related shift from rapidly rotating massive stars to low-mass stars that spin down under the influence of magnetized winds \citep{kratbreak1967,skumanichrotationlithium,2015somersmarc,kraftbreak2024}. Since the Kraft break ought to coincide with the lithium dip, we compute YREC-predicted $\tau_{CZ}$ for each field star at the main sequence turnoff. The convective overturn timescale $\tau_{CZ}$ is used in cool stars to predict stellar activity at fixed $\mathrm{P_{rot}}$. In Figure \ref{fig:taucztest}, we plot the median $\tau_{CZ}$(days) against the unscaled/YREC-predicted mass (top panel) and $\mscaled$ (bottom panel) for all metallicities. We compare these to the Kraft break computed in \citet{ngc7789lidip} and our scaled lithium dip range. The result yields excellent agreement between the YREC-predicted Kraft break for GALAH stars, the NGC7789 lithium dip mass/Kraft break, and the detection fraction dip range. We can also see that the stacked detection fraction, similar to the one computed for Figure \ref{fig:detec_frac_dip}, aligns with the transition between the predicted $\sim2$\,dex change in $\tau_{CZ}$. Interestingly, stars above $\feh\gtrsim0$ are predicted to spin down slightly slower, suggesting a somewhat broader transition at higher $\feh$. 

\begin{figure}
    \centering
    \includegraphics[width=1\linewidth]{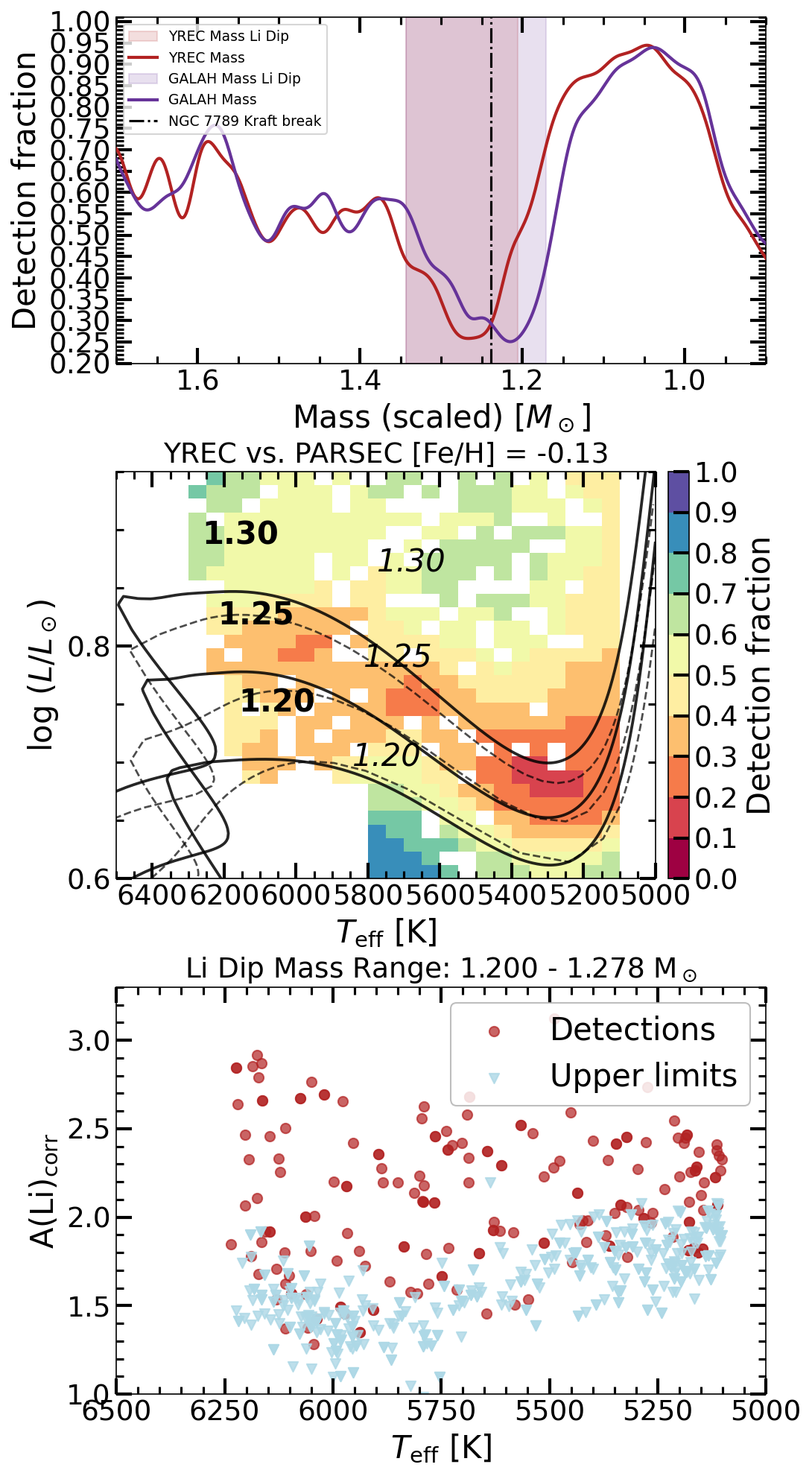}
    \caption{\textbf{TOP:} A 2D histogram depicting the lithium dip across the HR Diagram. The HR Diagram of our sample, colored by FDU-corrected lithium. There is a noticeable band of depletion that follows a stellar track corresponding to the center of the lithium dip of that metallicity in Table \ref{tab:lidip}. \textbf{BOTTOM} The Lithium dip band shown above for $\feh=-0.13$, plotting $\ali$ as a function of $\teff$.}
    \label{fig:stellartrack_lidip}
\end{figure}

This detection fraction in the dip's center raises the question: what else could it be? Systematic offsets in the extinction (see Section \ref{Dataset Reduction}) could translate to offsets in $\loglbol$ (and thus our YREC-computed masses). These offsets, if primarily detections, could scatter detections from the cool plateau or hot fork into the lithium dip. As a test of this, we compute the star-by-star mass offsets that would result from a difference between the Gaia 3D dustmap extinctions and the GALAH ones, and recompute the detection fraction KDE from the resulting offset masses. In doing so, we find the detection fraction in the dip actually \textit{increases}, ruling this explanation out.

Systematic errors in the mass interpolation across the subgiant branch could also blur such a sharp feature. We compare YREC with GALAH / PARSEC in the top two panels in Figure \ref{fig:stellartrack_lidip}. In the topmost panel, we compare an $\mscaled$ derived from $\mathrm{M_{GALAH}}$ using Equation \ref{equ:massscald} to the prior one from $\mathrm{M_{YREC}}$. The GALAH masses result in a dip center of 1.24$\msun$, which is closer to the cluster. The resulting width of $0.018\msun$ is $0.03$ larger than $\mathrm{M_{YREC}}$, but produces the same depth. Moreover, the middle panel shows a roughly uniform dip in detection fraction across the Hertzsprung gap, reaching a minimum but never falling to zero. \textit{There is no domain in the HR diagram where we see no detections.}  Both the YREC and PARSEC mass tracks trace the topology of the dip.

A final possibility is contamination from unidentified photometric binaries. Our contamination/background fraction is $\sim 32\%$; so a sample contamination at this level is not impossible. We would expect main sequence binaries - the most common contaminant - to be concentrated close to the hot side. However, the bottom panel of Figure \ref{fig:stellartrack_lidip} shows an even spread of detections and upper limits in $\teff$. Special conditions -  a large fraction of the sample with masses very close to being equal - would be needed for them to produce such a uniform detection pattern. Additionally, binaries found through RV variations should have been covered by our catalogs. While astrometric binaries could exist in this dataset, faint MS companions will not strongly impact $\loglbol$, and by extension mass.

The question now becomes: could the literature have seen detections in the lithium dip and simply overlooked them? Interestingly, \citet{paceclustm67} reported (but did not elaborate on) a small handful of detections in the M67 subgiants between ($\sim1.28-1.34\msun$) -- falling exactly within our $\feh_{\odot}$ lithium dip. We thus conclude that \textit{the detections in the lithium dip are real and warrant a further detailed analysis}. We discuss possibilities in Section \ref{discussion}.

\subsection{The Cool Side \& Cool Plateau Region} \label{coolplat_section}
\begin{figure*}
    \centering
    \includegraphics[width=1\linewidth]{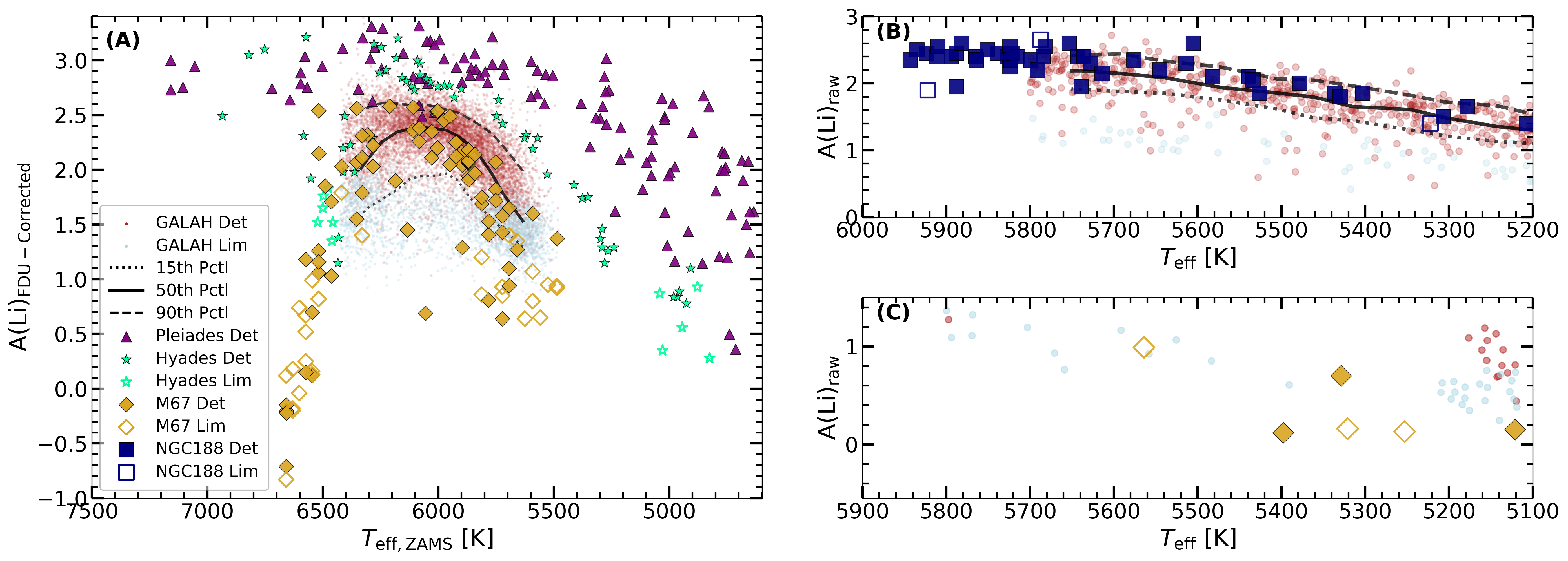}
    \caption{\textbf{\textit{(A)}}Cool plateau $\ali$ as a function of $\teff_{ZAMS}$. Red stars are GALAH detections, blue stars are upper limits, and isolated by taking stars within the mass range where the detection fraction is $\geq80\%$. The Hyades, Pleiades, and M67 are plotted against the GALAH-corrected lithium data (red and blue points). The 15th, 50th, and 90th rolling percentiles of the GALAH data are plotted as black lines. \textbf{\textit{(B)}} Subgiants from NGC 188 are plotted against their mass, metallicity, and HR diagram analogs in the GALAH data. The $\ali$ of the stars are plotted as a function of the normal $\teff$. The 15th, 50th, anf 90th rolling percentiles of the GALAH data are shown. \textbf{\textit{(C)}} The same as \textbf{\textit{(B)}}, but for subgiants in M67 isolated using isochrone fitting. Not percentiles are shown due to the low number of points.} 
    \label{fig:coolplat_slope_widths}
\end{figure*}

The cool plateau is the area of the lithium pattern with the richest literature \citep{zapaladepletioninfstars,spiteandspite,  marcrotmix1989, soderblom1995, sesrandmixing2005, Soutom67}. We define the cool plateau as the set of stars which have a mass below the lower mass bound of the lithium dip ($\mscaled=1.20\msun$), are younger than the field turnoff age of $\sim9$\,Gyr, and have a detection fraction of $\geq 80\%$. We compare the cluster and field patterns in Figure \ref{fig:coolplat_slope_widths}. However, the cluster stars are at varying stages of evolution. While M67 and NGC\,188 have subgiants, the younger Pleiades and Hyades do not. As investigators in the field tend to work in $\teff$ space, we therefore perform two calculations: we transfer the field stars to the younger cluster context, where $\teff$ is a mass proxy, and we use ISTARY isochrones to draw out analogs to our cluster subgiants. In the former case, we use mass, $\feh$, our YREC track suite, and \texttt{Kiauhoku}; we interpolate theoretical estimates of each star's temperature at the main sequence turnoff and zero-age main sequence. For the latter case, we match stars within $\feh\pm0.01 $, $\teff\pm10K$, and $\loglbol\pm0.01$ to isochrone analogs of the older clusters.

The resulting Figure \ref{fig:coolplat_slope_widths} shows the GALAH cool plateau detections (red), upper limits (blue) compared to the Pleiades, Hyades, M67, and NGC\,188 samples (purple, teal, gold, and dark blue, respectively). Panel (A) shows the whole GALAH cool plateau population at the $\teff$ they would have had at the zero-age main sequence. Here, several important features stand out. Lithium depletion increases with age across all masses, and shows dispersion at fixed $\teff$ in both the clusters and the field. The field abundances sit below the younger Hyades and Praesepe as a consequence of their age. The field abundances are close to, but slightly above, M67 -- raising the possibility that diffusion from \citet{Soutom67} depresses MS lithium relative to the turnoff. All of these features are generic predictions of models with rotationally induced mixing \citep{marcpapercosmoli,marcrotmix1989}. Panel (B) shows the rich NGC\,188 subgiant dataset and its GALAH field star analogs in normal $\teff$. The field stars exhibit more dispersion and upper limits, with the median abundance sitting slightly below ($\sim0.01$\,dex) the cluster. This is likely a systematic difference in the measurements. In Panel (C) we perform the same exercise compared to the much sparser M67 subgiant sample. Both show dispersion and similar abundances. 

\begin{table*}
\centering
\caption{Cool plateau A(lithium) summary by \feh\ bin. A(lithium) percentiles (15th, 50th, 90th) are
measured from the rolling-peak percentile structure in mass-scaled space with quoted uncertainties given by the median upper/lower observational errors
($e_{\rm ALi,upp}$, $e_{\rm ALi,low}$) of stars within each percentile's own rolling-bin
window. Mass values are the min/median/max of the native mass distribution across the pooled
detections in each \feh\ bin, with uncertainty given by the median hybrid mass error
($\sigma_{M,\rm hybrid}$) over that same pooled population. The quoted $\teff$ intervals give
the temperature ranges of stars lying within the A(lithium) rolling-bin window for the ZAMS and
MSTO mappings separately. Asterisks ($*$) denote values in bins where upper limits exceed
20\% of the stars, indicating the reported measurement may be influenced by non-detections.}
\label{tab:cool_plateau}
\hspace*{-80pt}
\begin{tabular}{l ccc ccc ccc ccc}
\hline\hline
 &
   \multicolumn{3}{c}{A(lithium)} &
   \multicolumn{3}{c}{Mass ($M_\odot$)} &
   \multicolumn{3}{c}{$T_{\rm eff,ZAMS}$ (K)} &
   \multicolumn{3}{c}{$T_{\rm eff,MSTO}$ (K)} \\
\cmidrule(lr){2-4}\cmidrule(lr){5-7}\cmidrule(lr){8-10}\cmidrule(lr){11-13}
\feh &
  $Pct_{15}$ & $Pct_{50}$ & $Pct_{90}$ &
  min & med & max &
  min & med & max &
  min & med & max \\
\hline
$-0.40$ & $1.37^{+0.042}_{-0.044}$* & $2.33^{+0.043}_{-0.044}$ & $2.55^{+0.043}_{-0.044}$ & $0.83\pm0.010$ & $0.93\pm0.010$ & $0.99\pm0.010$ & 5816 & 5996 & 6130 & 6045 & 6157 & 6246 \\
$-0.30$ & $1.97^{+0.042}_{-0.050}$* & $2.33^{+0.042}_{-0.050}$ & $2.54^{+0.042}_{-0.051}$ & $0.87\pm0.011$ & $0.98\pm0.011$ & $1.05\pm0.011$ & 5912 & 6037 & 6195 & 6080 & 6163 & 6274 \\
$-0.20$ & $1.95^{+0.041}_{-0.042}$* & $2.36^{+0.041}_{-0.042}$ & $2.57^{+0.041}_{-0.042}$ & $0.93\pm0.012$ & $1.02\pm0.012$ & $1.10\pm0.012$ & 6027 & 6107 & 6217 & 6143 & 6192 & 6242 \\
$-0.10$ & $2.06^{+0.042}_{-0.046}$* & $2.36^{+0.042}_{-0.046}$ & $2.55^{+0.041}_{-0.047}$ & $0.98\pm0.013$ & $1.07\pm0.013$ & $1.16\pm0.013$ & 5999 & 6089 & 6234 & 6098 & 6160 & 6231 \\
$+0.00$ & $2.13^{+0.037}_{-0.038}$* & $2.41^{+0.036}_{-0.038}$ & $2.57^{+0.036}_{-0.038}$ & $1.04\pm0.013$ & $1.12\pm0.013$ & $1.21\pm0.013$ & 5960 & 6092 & 6242 & 6061 & 6117 & 6209 \\
$+0.10$ & $2.14^{+0.031}_{-0.032}$* & $2.42^{+0.031}_{-0.032}$ & $2.64^{+0.033}_{-0.033}$ & $1.09\pm0.015$ & $1.17\pm0.015$ & $1.27\pm0.015$ & 5976 & 6093 & 6237 & 6062 & 6107 & 6179 \\
$+0.20$ & $1.95^{+0.027}_{-0.029}$* & $2.43^{+0.027}_{-0.029}$ & $2.63^{+0.027}_{-0.029}$ & $1.14\pm0.017$ & $1.22\pm0.017$ & $1.31\pm0.017$ & 5930 & 6055 & 6204 & 5997 & 6053 & 6131 \\
$+0.30$ & $1.72^{+0.025}_{-0.023}$* & $2.43^{+0.025}_{-0.023}$ & $2.72^{+0.033}_{-0.032}$ & $1.20\pm0.017$ & $1.27\pm0.017$ & $1.37\pm0.017$ & 5859 & 6084 & 6185 & 5958 & 6047 & 6097 \\
$+0.40$ & $1.50^{+0.060}_{-0.066}$* & $2.37^{+0.028}_{-0.025}$* & $2.80^{+0.031}_{-0.026}$ & $1.25\pm0.016$ & $1.29\pm0.016$ & $1.39\pm0.016$ & 5856 & 5934 & 6085 & 5916 & 5959 & 6039 \\
\hline
\end{tabular}
\end{table*}
Previously, the cool plateau domain was only extensively observed at near-solar metallicities. In contrast, our data allow us to infer the median lithium abundance, median mass, and width of the cool plateau across a range of metallicities. We summarize these findings in Table \ref{tab:cool_plateau}, which characterizes the cool plateau. We note that some of the 15th percentile values are upper limits. The mass shifts here are the same as the ones we inferred earlier for the lithium dip and the cool plateau (see Table \ref{tab:lidip} and Figure \ref{fig:detec_frac_dip}), so the two features are clearly astrophysically related. A striking feature of this data is that the absolute depletion in the cool stars is a remarkably weak function of metallicity. The median lithium ($\langle \ali\rangle=2.42$) does not deviate by more than $\sim0.1$ dex between metallicities. 

\begin{figure}
    \centering

    \includegraphics[width=1\linewidth]{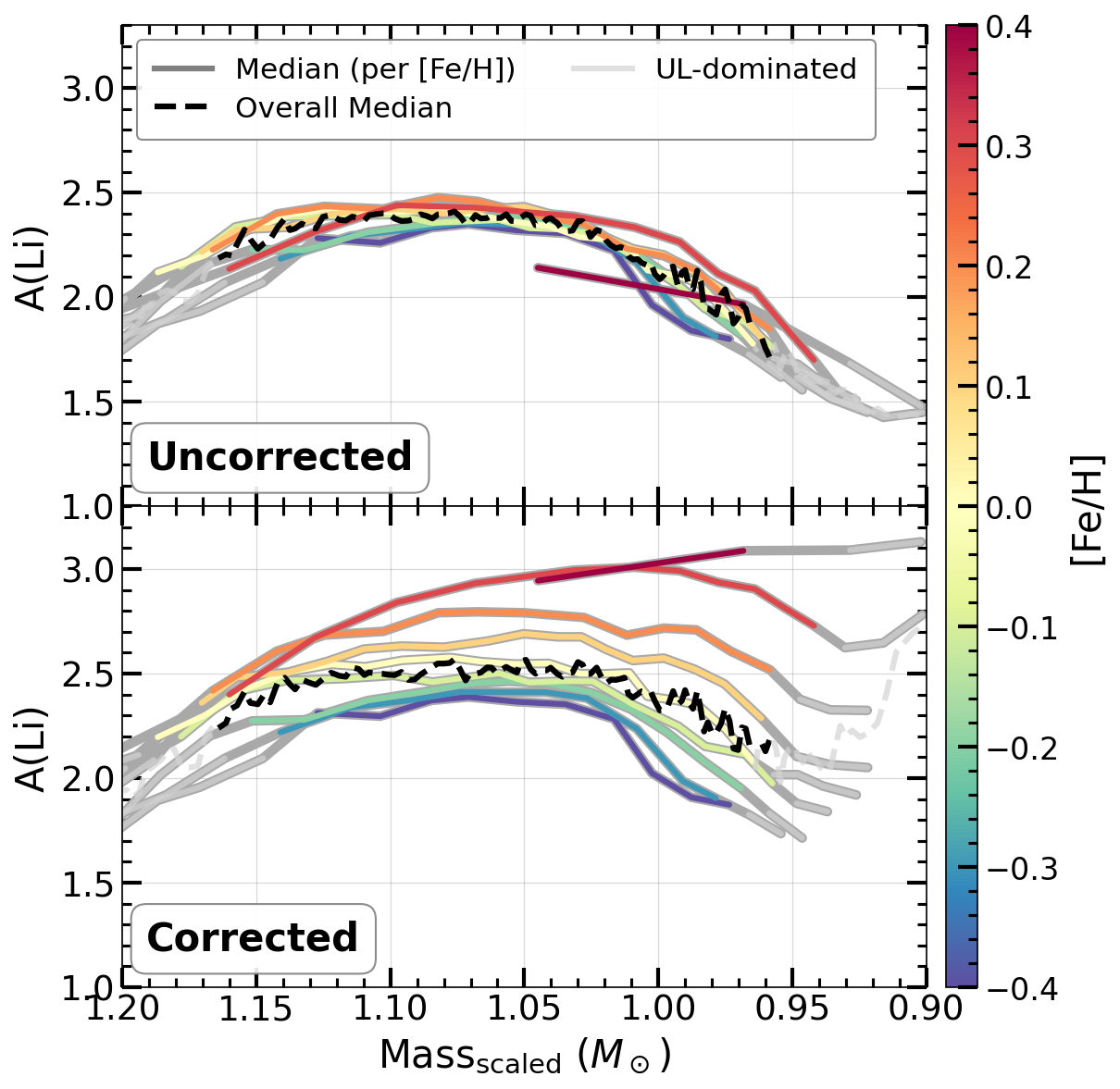}
    \caption{Before (\textbf{TOP}) and after (\textbf{BOTTOM}) the YREC pre-main sequence (PMS) correction, plotted against $\mscaled$, colored by $\feh$. The median trend of the whole sample is plotted as a dashed black line. All lines become grey when the upper limits begin to dominate (as in Figure \ref{fig:massdredge}. $\ali$ was pre-corrected for FDU depletion before correcting for PMS depletion. Lines are rolling medians of GALAH lithium data as a function of $\mscaled$ containing 200 stars per bin.}
    \label{fig:pmsslope}
\end{figure}

Although we are not in a position to reconstruct the \textit{birth} abundances directly, we can infer the end-of-main sequence data, which can be combined with a chemical evolution model to infer stellar depletion as a function of mass and metallicity. In addition to observational comparisons, theoretical ones can be made as well. Standard models predict a metallicity-dependent depletion trend for stars below the cool plateau. Close to solar metallicity, the strong mass trends are clearly seen in the Hyades and Pleiades, even in Figure \ref{fig:pmsslope}. The culprit is pre-MS depletion, mostly in the first 10\,Myr \citep{ibenpmsdepletion,bodenheimerpmsdepletin,bildstenpms}. Theoretical models predict lithium depletion increasing with $\feh$ and decreasing mass, as the accompanying increased opacity and slower contraction lead to a cooler envelope and deeper convection \citep{hayashi_baseofconvec,ibenpmsdepletion,bildstenpms, jeffriespmsli}. It is, however, an interesting question as to whether there is also mass-dependent mixing, and whether standard models predict the correct metallicity trends.

We use our YREC models to predict standard model pre-MS depletion. To do so, we compute the depletion factor between the deuterium-burning bithline and the main sequence, defined as the point where the central hydrogen mass fraction has dropped by $0.005$ from the birth value. We then add this factor to the dredge-up-corrected lithium of our sample below the lithium dip, which we show in Figure \ref{fig:pmsslope}. We find that mass-dependent and temperature depletion is apparent; the cooler, lower-mass pre-main sequence stars deplete more than stars near the Kraft Break ($\sim1.23\mscaled$) or the cool plateau peak ($1.05\leq\mscaled$). The trends are greyed out in regions where upper limits dominate ($\geq 50\%$ of stars), and depletion becomes severe. These mass and temperature trends in corrected models indicate main sequence depletion across all $\feh$. For stars $1.0\lesssim\mscaled\lesssim1.1$, pre-main sequence appears to extend the plateau over a wider range of masses, but not the overall levels. At $\feh>0$, however, the plateau extends and is even elevated as $\feh$ increases. This result is quite intriguing in the context of the cool plateau's observed constant $\ali$-$\feh$  (see Table \ref{tab:cool_plateau}). Thus, this elevation at higher $\feh$ raises numerous possibilities.

One such possibility is that systematics in the models or their input physics produce the elevated plateau at high $\feh$. However, our pre-main sequence depletion estimates are model-derived. The resulting errors would be difficult to distinguish from metallicity-dependent main sequence depletion or variations in initial lithium abundance. Testing this would require a systematic exploration of how the relevant input physics depends on metallicity, which is beyond the scope of this work.

Another explanation is that the models overcompensate for pre-main sequence depletion, and all stars deplete to the constant abundances we observe. Using Gaia ESO EWs, \citet{jeffriespmsli} indicated that true pre-main sequence metallicity dependence could be minimal.  However, these predictions from \citet{jeffriespmsli} only covered a near-solar metallicity range. A larger sample size of metal-rich clusters would be needed to fully determine this. 

The final avenue is that the depletion is real, and the metal-rich stars are born with more lithium. \citet{gao2020} also indicated an upward trend in $\ali$ at supersolar metallicities for GALAH cool plateau field stars, attributable to galactic production methods \citep{liproduction, lievolutionthinthickdisk}. Observations indicate a range of abundances in the solar neighborhood \citep{liproduction, lievolutionthinthickdisk, liradialmigration}, with the solar meteoritic values sitting at $\ali=3.3$\,dex \citep{andersgrevesemeteoritic}. Moreover, studies of the galactic bulge, thin and thick disk indicate that metal-rich stars could be born with higher lithium and migrate to other areas of the galaxy \citep{bulgeli,liradialmigration}. Thus, it is possible that at high metallicities, higher birth abundances are counterbalanced with increased depletion brought on by their higher opacities and deeper convective zones. If this scenario holds, then \textit{metal-rich stars that are older than the pre-main sequence, lie below the cool plateau peak, and retain a measurable lithium abundance must have been born with an elevated initial abundance}.

\subsection{The Hot Fork} \label{hotforksection}
\begin{deluxetable*}{lccccccccc}
\tablecaption{Detection and upper-limit statistics for the hot shelf, split by proximity to the lithium dip boundary. \label{tab:bluestragglers}}
\tablehead{
\colhead{} & \colhead{} & \colhead{} & \colhead{} & \multicolumn{3}{c}{Near Dip ($M < M_{\rm cut}$)} & \multicolumn{3}{c}{Far from Dip ($M > M_{\rm cut}$)} \\
\cline{5-7} \cline{8-10}
\colhead{[Fe/H]} & \colhead{$N$} & \colhead{$M_{\rm min}$} & \colhead{$M_{\rm cut}$} & \colhead{Det} & \colhead{UL} & \colhead{Det\%} & \colhead{Det} & \colhead{UL} & \colhead{Det\%}
}
\startdata
-0.40 & 504  & 1.20 & 1.50 & 168  & 185  & 47.6 & 37  & 114  & 24.5 \\
-0.30 & 759  & 1.25 & 1.55 & 343  & 251  & 57.7 & 56  & 109  & 33.9 \\
-0.20 & 1156 & 1.30 & 1.60 & 532  & 381  & 58.3 & 117 & 126  & 48.1 \\
-0.10 & 1241 & 1.35 & 1.65 & 487  & 419  & 53.8 & 181 & 154  & 54.0 \\
+0.00 & 1170 & 1.41 & 1.71 & 412  & 393  & 51.2 & 199 & 166  & 54.5 \\
+0.10 & 732  & 1.46 & 1.76 & 228  & 272  & 45.6 & 114 & 118  & 49.1 \\
+0.20 & 326  & 1.52 & 1.82 & 106  & 126  & 45.7 & 46  & 48   & 48.9 \\
+0.30 & 59   & 1.57 & 1.87 & 15   & 31   & 32.6 & 6   & 7    & 46.2 \\
+0.40 & 12   & 1.62 & 1.92 & 5    & 4    & 55.6 & 0   & 3    & 0.0 \\
\hline
Total & 5959 & 1.36 & 1.66 & 2384 & 2121 & 52.9 & 668 & 786  & 45.9 \\
\enddata\tablecomments{`Near Dip' includes stars within 0.3\,$M_\odot$ of the minimum mass in each \feh\ bin; `Far from Dip' includes stars above that mass threshold. Per-\feh\ rows use mass; the Total row uses $M_{\rm scaled}$.}
\end{deluxetable*}
For stars above the lithium dip, the much more abundant field star data present a strikingly different pattern than the clusters (see the top row panels of Figure \ref{fig:cluster_field_dip}). The range of field abundances is much larger than that seen in clusters, and while detections with high lithium abundance are present, they fall below the detections seen in the clusters. While a similar dispersion in clusters is seen just above the dip (within $0.3\msun$), the domain above it gives way to nearly all detections \citep{gao2020, clustwiynngc2056vsini, ngc7789lidip}. The field star data show only weak mass dependence, making it unlikely that mass errors or main sequence binaries are responsible for this dispersion.

The hot fork, true to its namesake, diverges into a minority of lithium-rich stars, mirroring the open clusters, and a majority of lithium-depleted stars. The top 10\% of Figure \ref{fig:massdredge} maintains a $\ali = 2.7$ with mass, and matches the cluster literature the closest. The bottom 50\% straddles the line between detection and upper limit domains at $\ali = 2.0$, with the lowest 15\% of our sample predominantly being ($\geq 50\%$) upper limits with abundances of $\ali\lesssim1.5$. In the metal-poor and metal-rich regimes, the fraction of upper limits and the upper limit boundary itself rises, leading to the detection fraction dropping to $ \leq 40\%$ (see \ref{fig:massdredge} \& \ref{fig:3x3stackedhistplot}). The broad range of depletion and abundances is universal across the masses and metallicities in this domain.

Several possibilities for this pattern stand out. Firstly, it is possible our data or methodology is greatly flawed. However, removing this dispersion requires us to abandon nearly 2/3rds of our sample. Our contamination/background fraction is nonzero ($\sim32\%$), so unremoved binaries are a possibility. Additionally, mass errors could scatter in depleted lithium dip stars. While the depletion and dispersion in the metal-rich regimes could be due to evolutionary effects, as many stars hotter than the lithium dip are not yet subgiants at our age cutoff, the metal-poor regimes may be a different story. \cite{bluestragglerslithium,marccrimlideplete} emphasize that blue straggler (BS) stars efficiently mix and deplete lithium during their merger formation process. While BS stars are often dwarfed by their MS compatriots in large field samples, the evolutionary timescales dictate there should be little to no high-mass stars above the dip at low and high metallicities. Given that we see a larger number of stars, some of them (especially those farther from the dip) must be blue stragglers. 

We construct Table \ref{tab:bluestragglers} to assess the possible population effects. We want to quantify the effects of contaminants, stars scattering in from the lithium dip, and also estimate the fraction of blue stragglers by comparing the statistics of metal-poor stars to the rest of the sample. We plot the number of detections and upper limits near the dip ($\leq M_{dip} +0.3\msun$), and far above it ($\geq M_{dip} +0.3\msun$). We include stars with $\feh\geq+0.2$ for reference, but their numbers are small due to evolutionary effects. 

The detection fractions are roughly constant across bins with two notable exceptions: the highest mass and lowest metallicity bins. Statistics are unsurprisingly poor in the most metal-rich bins. However, the upper limits in the metal-poor bins are likely populated by blue stragglers that crossed into the SGB. From these stars, we estimate that roughly $\sim1/3$rd of these upper limits could be blue stragglers. Some of the remaining depletion could come from dip stars that have scattered upward in mass. However, this is unlikely to explain all of the remaining stars. This is especially true for those $\geq 0.3\,\msun$ above the dip, which are too far from the dip to be easily explained by scatter. 

Alternatively, the disparity between the clusters and field stars could be errors in the cluster data or physical effects. The stars above the dip in \cite{ngc7789lidip} are very hot and generally rapidly rotating -- blurring both the Li line and the rotation measurements. It is possible, then, for these to be marginal detections. We therefore suggest that a re-analysis of the turnoff cluster measurements could be beneficial. At the same time, we have no direct evidence that the measurements are flawed, which leads us to the third possibility: the mapping between the cluster snd field may be the issue. 

The lack of strong trends with mass, unlike the stars below the plateau in Figure \ref{fig:massdredge}, could represent multiple physical effects mediated by rotation and levitation effects. Stars above the Kraft break have a wide range of rotation rates \citep{kratbreak1967, kraftbreak2024}, and therefore differences in the degree of rotationally induced mixing could produce the dispersion we see. The dispersion could also be enhanced by threshold effects. Mixing is known to be inhibited by mean molecular weight gradients, which are naturally produced by gravitational settling. We might therefore expect mixing to be generically suppressed in chemically peculiar stars on the main sequence. It is also the case that slow rotation is correlated with the onset of surface abundance anomalies. Given the fraction of chemically peculiar stars, lithium should be depleted in all but $\sim$1/3 of stars, consistent with our observations \citep{Abt_2017}. This mechanism could explain why lithium survives only in a minority of these stars \citep{Michaud1970, Abt_2017}. 

In addition, radiative levitation concentrates lithium in a thin layer near the surface. If it is operating in these hot stars, this would imply that lithium should dilute much faster than predicted by standard models when the "normal" subsurface layers are mixed in \citep{Michaud1970,abt1979, Abt1995, Abt_2017}. 

Unfortunately, we cannot directly test whether a range of rotation rates is producing the lithium dispersion for our sample; hot subgiants are rapidly spun down on the subgiant branch \citep{vansadersyrec2012}, and any pre-existing range of rotation rates would be erased. The first dredge-up would also erase abundance anomalies. However, it may be useful for future work to focus on comparing lithium in chemically peculiar and normal stars on the MS, which could test our theory.

\subsection{Comparison of Findings with Previous Field Data} \label{usand gaocompare}
\begin{figure}
    \centering
    \includegraphics[width=1\linewidth]{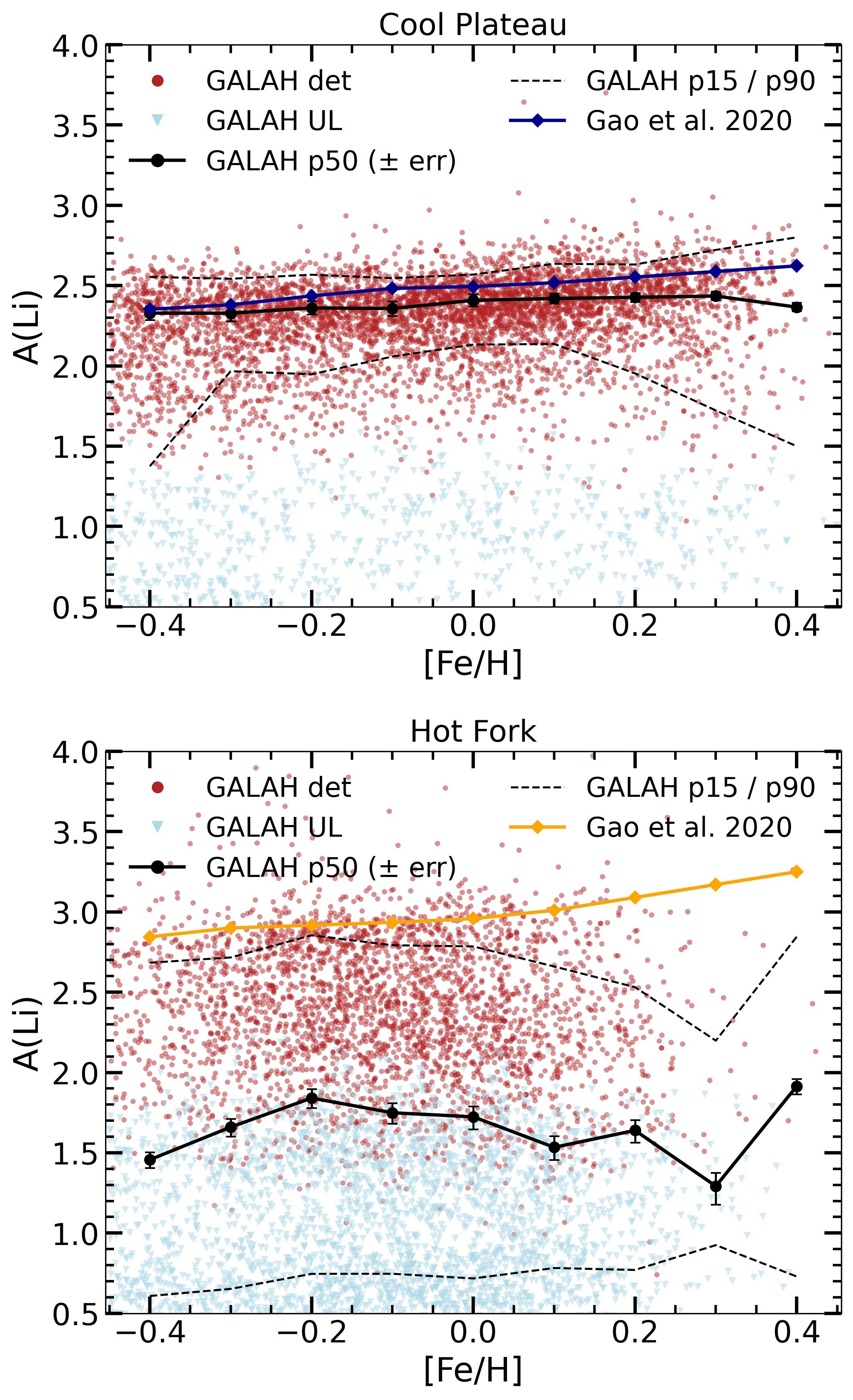}
    \caption{Cool Plateau (TOP) Hot Fork (BOTTOM) A(Li) as a function of \feh. We compute the rolling 15th, 50th, and 90th percentiles using \textbf{only detections} every 200 stars, shown as the black solid and dashed lines, in a similar fashion to \cite{gao2020}. Detections are in red and upper limits are in blue. The 50th percentile remains nearly constant at A(lithium) $\approx 2.4$ across the full \feh, while the 15th--90th spread grows from $\sim$0.3\,dex at \feh $= -0.3$ to $\sim$0.9\,dex at \feh $= +0.4$ due to small number statistics and population effects. We show the trends from \citet{gao2020} for reference.}
    \label{fig:gao_vs_us}
\end{figure}
In previous sections, we used cluster data to guide and contextualize our novel findings due to their rich history and well-founded theoretical predictions. With the data well understood, we compare it here to its progenitors found in \cite{gao2020}'s recent field studies of GALAH main sequence to early subgiant stars. The work particularly focused on the identification of a lithium abundance plateau in $\feh$ in both stars cooler and hotter than the lithium dip, providing an excellent comparison point.

Figure \ref{fig:gao_vs_us} shows the $\ali$ vs $\feh$ cool plateau (top) and hot fork, compared to the trends from Figure 2 of \cite{gao2020}. The median abundances (50th percentile) of our subgiants are $\sim0.1$\,dex lower on the cool plateau and $\sim0.5$\,dex lower on the field hot fork. These differences have a straightforward explanation:  \cite{gao2020} reported the median values for \textit{detections}, while we \textit{also} consider the numerous upper limits. The upper envelope of the data is higher for the hot stars, which was a key argument in  \cite{gao2020}. However, \textit{we differ by noting that there is a large dispersion}. Li may be preserved in some hot stars, but it is definitely destroyed in many. 

\section{Predicting the Lithium of the RGB} \label{lirichgiants}
\begin{figure*}
    \centering
    \includegraphics[width=1\linewidth]{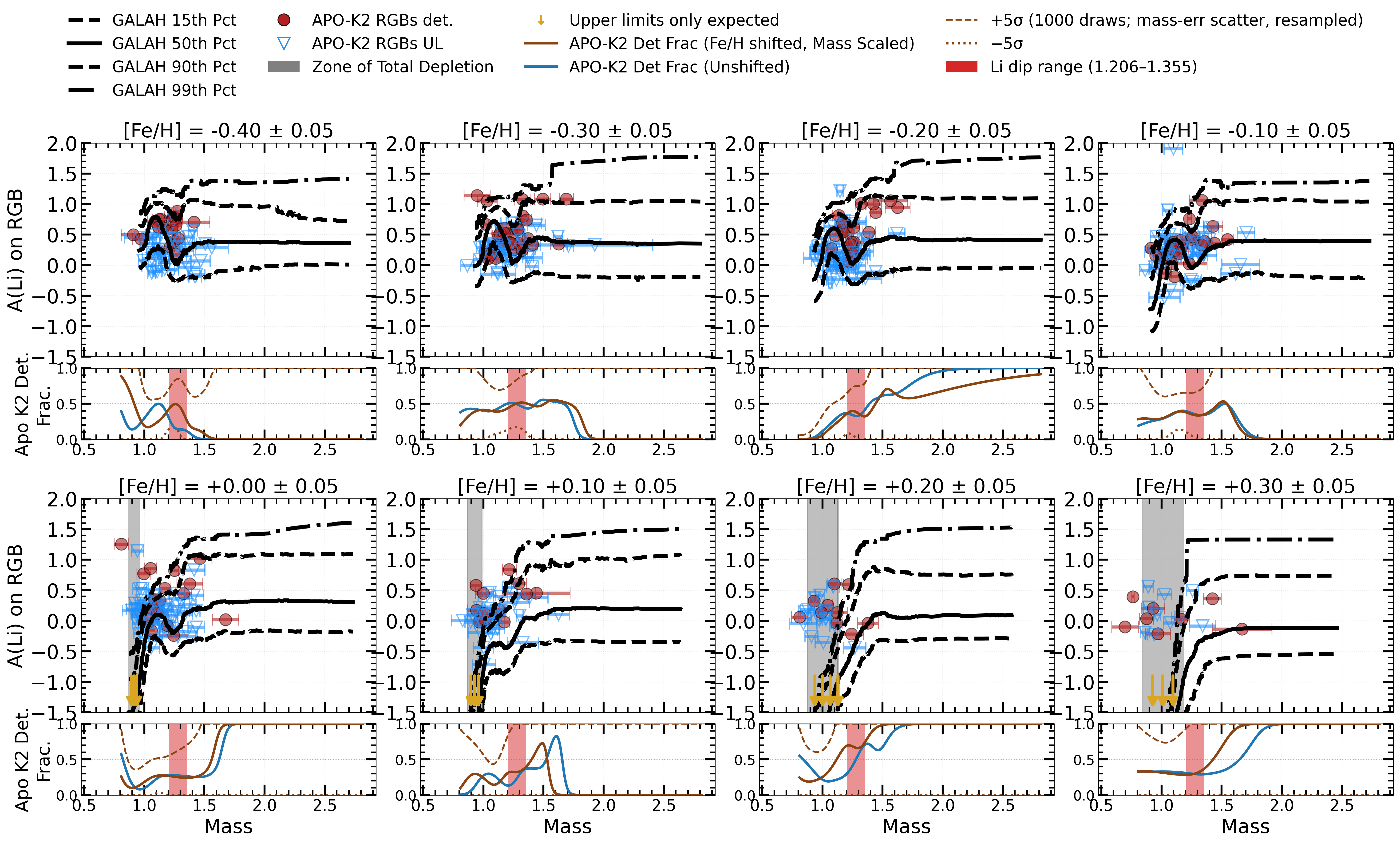}
    \caption{Lithium predicted for GALAH stars on the RGB against YREC mass, compared to the  \citet{apok2cat} RGB stars. The black lines represent the 15th, 50th, and 90th percentiles of the GALAH $\ali$ projected onto the lower RGB. The APO-K2 RGB detections (red circles) and upper limits (blue triangles) are shown alongside their asteroseismic mass errors. The detection fraction by mass for the APO-K2 RGB stars is shown below each plot. The blue line represents the raw KDE of the detection fraction, whereas the brown line indicates the fraction after the $\feh$ and mass binning effects were accounted for (same as the method in Figure \ref{fig:cluster_field_dip}). The dashed and dotted brown lines indicate the $\pm5\sigma$ of a synthetic population created by sampling stars 1000 times in a Monte Carlo simulation from a uniform distribution built on each star's $\sigma_{\text{Apo-K2 Mass Error}}$. Golden arrows and grey shading indicate the regime where the models predict total destruction of Li on the RGB.}
    \label{fig:lirichgiants}
\end{figure*}

We have demonstrated that our subgiant data is a natural extension of main sequence trends. We now perform the logical next step of comparing our results with those of more evolved stars. First, we can check whether our data are consistent with trends predicted by the full first dredge-up. Second, this may shed light on the origin of lithium-rich giants. For red giants, dredge-up strongly dilutes lithium \citep{iben1967fdupostms_b,iben1974review,kailimarc2025}; despite this, there is a small population with high lithium \citep{galahrgblistudy,tayr_lirich_rgh_limit, delarezalirichrgb}. In some cases, post-MS production is absolutely required \citep{cameronfowler, liproduction}. However, the exact threshold is controversial, as some stars leave the MS with high abundances \citep{chanamemarc}; therefore, some lithium-rich stars can be explained without lithium production or ingestion. Other stars have lithium destroyed, not diluted, during dredge-up \citep{kailimarc2025}; even a low level of detected lithium requires production for these stars. The lithium dip should also propagate into the giant domain, which might be another domain where detections likely require production \citep{claudiagomezengulfment}.

We use our models to predict both the main sequence turnoff value and the lower red giant branch value for any given star. For stars below the red giant branch bump, in situ mixing is not expected or seen even in systems with strong upper giant branch destruction \citep{lindmixingngc6397}. To link our subgiant projections to observed giants, we use GALAH stars with asteroseismic masses from the APO-K2 catalog \citep{apok2cat}, including only stars classified as being on the lower red giant branch; see \citet{kailimarc2025} for a discussion. This gives us a sample of 570 stars. We deplete the FDU-corrected lithium by the difference between the turnoff lithium and the lithium at the median luminosity on the lower RGB domain\footnote{We define the lower RGB as $2.5\leq \logg \leq 3.6$ and $\teff \leq 5100$~K.} for each star's mass and metallicity. Figure~\ref{fig:lirichgiants} shows the results. For subgiants, the lithium that we predict for the lower RGB is plotted as a function of YREC mass, where they are compared to the APO-K2 RGB stars. The gray bands represent mass and metallicity domains where depletion is projected to drive abundances to an undetectable level, $\ali<-1.5$.

Our projected data show similar trends to those seen in \citet{chanamemarc}, but use a much more precise method. For most stars, the detections bracket the observed range, and the upper limits are consistent with it. However, in descendants of stars from below the cool plateau, the models predict a forbidden zone where no detections should be seen. This is due to a combination of a low turnoff abundance, dilution, and destruction during dredge-up \citep{iben1967fdupostms_b,iben1974review,kailimarc2025}. Detections in this domain require either lithium production, masses strongly impacted by interactions, or a severe breakdown in models of the FDU. These trends are largely confirmed at low metallicity, where we predict $\ali\leq -0.5$. 

Interestingly, APO-K2 appears to indicate a median detection fraction of $0.286$ in the forbidden zone at $\feh\geq0.0$, where standard models predict ferocious rates of destruction (shaded grey mass range). Of the stars in $\feh \geq 0.0$, 48 of them (33 upper limits and 15 detections) fall within this mass range. These could be higher-mass stars, where detections are expected, scattering into the forbidden zone due to errors in their inferred mass. Stars could also scatter in between $\feh$ bins, appearing to land within a mass and metallicity domain of destruction. We therefore test this claim. We assume a Gaussian probability distribution for each APO-K2 star given its mass and $\feh$ error. We perform a 1000 draw Monte-Carlo simulation to produce a synthetic distribution. From this, we calculate a rolling detection fraction, yielding $0.273\pm0.038$. Thus, we can assume measurement errors can account for the detections we see.

It is important, though, not to discount explanations such as errors in the models. Models on the RGB often fail to reproduce physical processes such as the Red Giant Bump without envelope overshooting -- which we exclude. However, \citet{kailimarc2025} indicated that while convenient, this solution moves the lithium destruction boundary to higher masses in contrast with the GALAH data. 

There is a chance the detections could also be a real phenomenon. \citep{galahlirichgiants} found that Li rich giants in GALAH DR2 have a statistically higher occurrence rate in metallicities above solar. Moreover, they found a connection between fast rotation and high lithium. This implicates stellar interactions, as typical giants are slow rotators. \citet{jamielirichgiants} corroborated this connection between Li rich giants and rotation, also indicating these stars exhibited chemical peculiarity that could result from engulfment. Both speculate that production mechanisms such as \citet{cameronfowler} are a final possibility. Thus, it is also likely that \textit {these represent an enhanced production or higher birth abundance in metal-rich giants}. 

Additionally, while the signature of the Lithium dip appears in our projected abundances, the detection fraction (computed in a similar manner to Section \ref{lidip_section}) lacks any obvious dip. As discussed in Section \ref{lidip_section}, this feature is easily smeared by mass errors and binning effects due to $\feh$ and mass. As a result, we perform a similar exercise to Figure \ref{fig:cluster_field_dip}, wherein we plot the stars in the $\mscaled$ coordinate to remove $\feh$ binning effects, and then plot the detection fraction curves resulting from the $\pm5\sigma$ of a synthetic population created by the same Monte Carlo simulation as before. However, unlike in the GALAH subgiant Li dip, while the method lowers the detection fraction in the dip by $\sim0.15$ at most, it does not resolve the dip. K2 astroseismic mass errors are larger than the ones we computed for the GALAH subgiants; therefore, it is possible that they still blur this feature out.

Turning to more massive stars, the lithium data span a wide range consistent with subgiant projections. This is fully consistent with the broad range that we see at the turnoff. For stars above the Kraft break, we predict RGB abundances $\ali \geq 1.5$~dex; \textit{this is striking evidence that much of the observed lithium-rich RGB abundance and dispersion can arise from standard FDU alone}. However, we also predict a minority of GALAH stars with high abundances even in the absence of production in our models. A small fraction of our projected stars ($\leq 10\%$) are predicted to reach near- or super-meteoritic abundances ($3.3\leq\ali\leq4.5$) \citep{liproduction, lievolutionthinthickdisk, liradialmigration}. We therefore caution that such stars need not be interpreted as evidence for lithium production.

\section{Discussion} \label{discussion}

We derive turnoff-accurate lithium abundances for 22,442 GALAH subgiants spanning $0.9\lesssim\msun\lesssim2.0$ and $-0.4\leq\feh\leq+0.4$, correcting for post-main sequence dilution from the first dredge-up. Our sample notably includes intermediate-mass stars above the Kraft break, where main sequence lithium measurements are compromised by rapid rotation and weak spectral features at high $\teff$ ($\gtrsim7000$~K).

We recover a lithium dip that persists across all metallicities in our sample. It is characterized by a strong, \textit{but nonzero}, local minimum in the detection fraction. The telltale feature threads a tight channel in luminosity, and therefore mass, across the HR Diagram. The mass regime coincides with the transition between radiative and convective envelopes, and the empirical metallicity trends match theoretically predicted ones for the radiative/convective boundary. \textit{Notably, the field lithium dip is not empty}: we observe a persistent population of stars with Li detections, of order 25\%, across all metallicities. Although many clusters have no such stars in the dip, there are exceptions analogous to the small handful found in M67 subgiants \citep{paceclustm67}. There could be a number of explanations. One family is mundane: these are actually lower-mass binary stars incorrectly assigned Li dip masses, or main sequence interlopers incorrectly considered as subgiants. We see these stars across a range of $\teff$, so main sequence contamination is unlikely. However, we cannot rule out subgiant and main sequence turnoff binaries as a possible source. Precise radial velocity studies with upcoming spectroscopic surveys or Bayesian HR-diagram placement using Gaia could reveal unresolved binaries \citep{msmsbinarycat}; binaries within GALAH's 2$''$ pixel could also masquerade as single stars, so cross-matching against Gaia DR4 at that pixel scale may further identify contaminants. Engulfment or interactions with a substellar companion provide additional possibilities. \citet{claudiagomezengulfment} showed that substellar-companion or engulfment could enrich lithium at frequencies consistent with our dip detection rate; this is appealing since engulfed stars would appear as ordinary single stars in most parameters and evade binary catalogs. \citet{engulfment_galah} similarly found that lithium enrichment from planet engulfment was extraordinarily detectable in subgiants over a narrow mass range. Confirming these, though, would require analyzing abundance trends, radial velocities, and condensation temperature \citep{lucyalpharichstars} --- beyond our present scope. 

Above the dip, we observe a very wide range of lithium abundances, with lithium detected in roughly half of the stars and a minority of stars at high abundances. The upper limits are in many cases quite stringent, indicating a very wide range of lithium depletion rates on the main sequence. We find that population effects, background contaminants, and blue straggler merger products scattered in by radial migration \citep{bluestragglerslithium,marccrimlideplete} can account for roughly $30-40\%$ of the non-detections. We also note that, even in the clusters, there is a large abundance scatter seen in stars just above the dip \citep{ngc7789lidip}. For the small number of stars at high abundances, we find these remain several tenths of a dex below analogs in open clusters. This difference raises three families of possibilities. This could reflect systematics in the cluster data, the field data, or in the mapping between the two. 

For the subgiant data, unrecognized binaries are the largest issue, which could be addressed with the upcoming Gaia DR4 release. For the cluster data, it may be useful to reanalyze the data and to distinguish between chemically peculiar and normal stars for the Li data. 

An interesting alternative, which we favor, is related to the interactions between radiative levitation, gravitational settling, and rotationally induced mixing. For slow rotation, gravitational settling induces $\mu$-gradients that suppress mixing. These slowly rotating A stars are in general chemically peculiar \citep{Abt1995}, so they could preserve lithium in a thin envelope near the surface \citep{Michaud1970,abt1979,michauddiffusionfstars,chaboyerrimm67gkstars,Abt1995,Abt_2017}. Furthermore, radiative levitation can concentrate Li in the surface layers; this would be diluted once stars evolved to the subgiant branch, explaining the offset between the higher MS values and the lower ceiling seen in the subgiants. In this picture, the more highly depleted stars are the ones where mixing overcame diffusion. In this case, the wide Li dispersion is a smoking gun for rotationally induced mixing, amplified by the suppression of mixing in chemically peculiar stars. We cannot test the correlation with rotation directly in the subgiants, as they spin down efficiently after they leave the main sequence and ignite magnetized winds. However, future modeling efforts and follow-up rotation measurements may be able to pinpoint regimes where rotation information is preserved enough for us to see a signal.

The cool plateau consists almost entirely of detections, and exhibits universal depletion that depends little on metallicity. Dispersion at a given mass is present, but smaller than for massive stars. The median abundances are lower than younger open-cluster analogs (NGC~7789, Pleiades) \citep{pleiadesbouvier,ngc7789lidip}, are comparable to the field peak in older ones (M67, NGC~188) \citep{gao_m67paper, ngc188rotmixlidip}, and we therefore view the cool plateau as an analog of the Spite plateau abundance \citep{spiteandspite, gao2020} observed in metal-poor stars. The plateau's location shifts in mass, but the mass difference between the lithium dip and the cool plateau is invariant. We can therefore explain the metallicity dependence as arising from the change in location of the Kraft break \citep{kratbreak1967,marcrotmix1989,marcpapercosmoli,marccrimlideplete}. 

Pre-MS depletion has a strong impact on stars cooler than the plateau. However, for stars above $\feh\geq+0.1$, pre-MS depletion is predicted to impact the mean value in the plateau itself. If main sequence mixing does not drop in metal-rich stars - and there is no obvious physical reason why it should do so - the most natural explanation is higher lithium birth abundances in metal-rich stars. The solar neighborhood hosts a wide range of lithium abundances, with meteoritic values sitting far above our cool plateau at $\ali=3.3$ dex \citep{andersgrevesemeteoritic}. This shows strong evidence of a balance between mass-dependent main sequence depletion and $\feh$-driven pre-main sequence depletion to values that depend only weakly on metallicity by the turnoff. Observations of young stars at higher metallicity close to the galactic center rather than the solar circle could confirm this. 

Our subgiants sample is a valuable starting point for projecting lithium backward (on the main sequence) and forward (the RGB and RC) in time. Combined with stellar models of various mixing prescriptions, these stars can be used to simulate the effects of various physical processes. We showcase this potential by projecting our turnoff abundances to the lower red giant branch using the lithium depletion from our standard models, comparing to APO-K2 RGB stars. While lithium abundance on the RGB is seen as a function of mass, it is generally considered that lithium abundances higher than $\ali\gtrsim1.5$ are anomalous \citep{galahrgblistudy,tayr_lirich_rgh_limit, delarezalirichrgb}. This remains true for our projected data; the upper envelope (upper 10\%) of our sample is of order $\ali\approx1.0$ dex. There is a population of stars (upper $1-2\%$) from above the Kraft break, for which we predict lithium abundances of up to order $ \ali\approx1.5$ dex can be achieved without production. 

Given that lithium is severely depleted in the dip, and further diluted, our models predict that the RGB should inherit the imprint of the lithium dip. However, we do not see an obvious Li dip signature in the APO-K2 stars. Based on limited testing, this may simply be a case of mass errors smearing out a feature that applies to a narrow mass range. More precise asteroseismic data may be needed to see this effect; it would be surprising if it were truly absent. Engulfment of a planet or substellar companion is also a plausible explanation for the near- and super-meteoritic abundances predicted for the RGB sample. 

On an interesting note, we observe a small population of APO-K2 stars with $\ali$ up to $\approx0.5$\,dex in the domain below the metal-rich ($\feh \geq 0.0$) cool plateau where our models predict there ought to be total destruction of lithium. A synthetic distribution of APO-K2 stars indicated this could arise from the random statistical scatter in mass and $\feh$. Despite this, it is still possible that some could be genuine products of lithium production and enrichment. An analysis of clusters with metal-rich RGB analogs could indicate if this is indeed occurring. 

Future missions such as Gaia DR4 will add thousands of radial-velocity and astrometric measurements that may clarify the dip detections, while GALAH Phase~2 will refine abundances across the existing catalog. Coupling these measurements together could refine the mass estimates and provide MS lithium measurements that could test whether slow rotators are chemically peculiar. In the same vein, we can yield ages for the subgiant catalog --- which, paired with GALAH Phase~2, could aid chemical age-dating via $\mathrm{[\alpha/Fe]}$, $s$-, and $r$-process elements. In this ever-expanding age of big astronomical datasets, our turnoff abundances could offer a strong future calibration benchmark. In a similar vein. In addition to the giants, our turnoff abundances could anchor the end conditions of a rotational mixing model, and subsequently de-evolve stars to recover the zero-age main sequence lithium pattern. 

\section{Acknowledgments}
MHP and VAS acknowledge support from NASA grant 80NSSC24K0637. Many thanks to Jennifer Johnson for her insightful review and comments. Thank you to Maddie Howell, Lucy Lu, Alexa Beyer, Jamie Tayer, and Jess Stasik for their helpful discussions. We also thank Sarah Martell and Sven Buder of the GALAH collaboration for their insights into the GALAH dataset. 

\newpage
\bibliography{references}{}
\bibliographystyle{aasjournal}

\section{Appendix} \label{appendix}

\input{table1_print.tex}

\input{table2_print.tex}

\input{table3_print.tex}

\input{table4_print.tex}

\input{table5_print.tex}

\end{document}

%% file: table1_print.tex
% 10-row excerpt
% Full table: 98,345 rows x 11 columns
% Print columns: 11

\begin{deluxetable*}{ccccccccccc}
\tablecaption{Stars Removed as Contaminants: Red Clump Stars, Sub-Turnoff Main-Sequence Binaries, and Catalog Binaries\label{tab:contaminants}}
\tablecolumns{11}
\tablehead{
\colhead{GALAH ID} & \colhead{Gaia DR3 ID} & \colhead{2MASS ID} & \colhead{f$_{\rm RC}$} & \colhead{f$_{\rm subTO}$} & \colhead{f$_{\rm MSMS}$} & \colhead{f$_{\rm Nataf}$} & \colhead{f$_{\rm Gaia,bin}$} & \colhead{$N_{\rm bin}$} & \colhead{f$_{\rm bin,cat}$} & \colhead{f$_{\rm contam}$}
}
\startdata
230805002001195 & 6726128075021752704 & 18222290-4138438 & 1 & 0 & 0 & 0 & 0 & 0 & 0 & 1 \\
170110001601118 & 5483849943301538944 & 07003105-5747219 & 0 & 0 & 1 & 0 & 0 & 1 & 1 & 0 \\
170713002101108 & 6105651226538287488 & 14441979-3810529 & 0 & 0 & 1 & 0 & 0 & 1 & 1 & 0 \\
170912001901025 & 6888510132645274752 & 21023976-1259330 & 0 & 0 & 1 & 0 & 0 & 1 & 1 & 0 \\
150405002001198 & 5631288478292936320 & 09211729-3156132 & 0 & 0 & 1 & 0 & 0 & 1 & 1 & 0 \\
150429004101161 & 6043491573615819392 & 16095491-2555405 & 0 & 1 & 1 & 0 & 0 & 1 & 1 & 1 \\
160123001601105 & 5561815389413488512 & 07193293-3954097 & 0 & 0 & 1 & 0 & 0 & 1 & 1 & 0 \\
160108002001276 & 67104706472021632 & 03560337+2532386 & 0 & 1 & 0 & 0 & 0 & 0 & 0 & 1 \\
150703004101389 & 5918200432625963904 & 17503664-5805015 & 0 & 0 & 1 & 0 & 0 & 1 & 1 & 0 \\
140814005401386 & 6381429908380652416 & 23100281-7043293 & 0 & 0 & 1 & 0 & 0 & 1 & 1 & 0 \\
\enddata
\tablecomments{This table is published in its entirety in machine-readable format. A portion is shown here for guidance regarding its form and content. The full table contains 98,345 rows and 11 columns.}
\end{deluxetable*}

% ── Column key ──
\begin{deluxetable}{lll}
\tablecaption{Column descriptions for Table~\ref{tab:contaminants}\label{tab:contaminants_cols}}
\tablehead{
\colhead{Column} & \colhead{Label} & \colhead{Description}
}
\startdata
\texttt{galah\_id} & GALAH ID & GALAH source identifier \\
\texttt{gaiadr3\_source\_id} & Gaia DR3 ID & Gaia DR3 source identifier \\
\texttt{tmass\_id} & 2MASS ID & 2MASS source identifier \\
\texttt{flag\_rc} & f$_{\rm RC}$ & Red clump flag \\
\texttt{flag\_subturnoff} & f$_{\rm subTO}$ & Sub-turnoff main-sequence flag \\
\texttt{flag\_msmscat\_binary} & f$_{\rm MSMS}$ & MSMS catalog binary flag \\
\texttt{flag\_natafcat\_binary} & f$_{\rm Nataf}$ & Nataf catalog binary flag \\
\texttt{flag\_gaia\_binary} & f$_{\rm Gaia,bin}$ & Gaia NSS binary flag \\
\texttt{n\_binary\_flags} & $N_{\rm bin}$ & Number of binary flags triggered \\
\texttt{flag\_catalog\_binary} & f$_{\rm bin,cat}$ & Unified catalog binary flag \\
\texttt{flag\_contaminant} & f$_{\rm contam}$ & Unified RC / sub-turnoff contaminant flag \\
\enddata
\end{deluxetable}

%% file: table2_print.tex
% 10-row excerpt
% Full table: 22,442 rows x 13 columns
% Print columns: 13

\begin{deluxetable*}{ccccccccccccc}
\tabletypesize{\scriptsize}
\tablecaption{Conservative Sample Stellar Parameters with YREC-Mapped Temperatures and Pre-Main-Sequence Lithium Parameters\label{tab:pms_undep}}
\tablecolumns{13}
\tablehead{
\colhead{GALAH ID} & \colhead{$T_{\rm eff}$} & \colhead{$\sigma_{T_{\rm eff}}$} & \colhead{$\log g$} & \colhead{$\sigma_{\log g}$} & \colhead{[Fe/H]} & \colhead{$\sigma_{\rm [Fe/H]}$} & \colhead{$\log(L/L_\odot)$} & \colhead{$M$} & \colhead{Age} & \colhead{$T_{\rm eff,ZAMS}$} & \colhead{$T_{\rm eff,MSTO}$} & \colhead{pms\_undep}
}
\startdata
160519002601288 & 5101 & 72 & 3.731 & 0.095 & -0.007 & 0.059 & 0.513 & 1.120 & 8.55 & 6039 & 6100 & 2.344676381760745 \\
150606001901019 & 5277 & 77 & 3.602 & 0.107 & -0.431 & 0.064 & 0.737 & 1.164 & 5.19 & 6665 & 6584 & 2.2082426937871915 \\
170415002001297 & 5575 & 68 & 3.963 & 0.102 & 0.357 & 0.055 & 0.477 & 1.188 & 6.84 & 5758 & 5882 & 3.1740210399920192 \\
150606004901094 & 5173 & 75 & 3.693 & 0.103 & -0.212 & 0.062 & 0.574 & 1.074 & 7.85 & 6173 & 6225 & 2.248791799107848 \\
170510007301313 & 6150 & 69 & 3.852 & 0.115 & -0.401 & 0.067 & 0.764 & 1.148 & 4.50 & 6614 & 6549 & 1.6198934543425734 \\
230203002101025 & 5728 & 69 & 3.013 & 0.104 & -0.084 & 0.059 & 1.774 & 2.404 & 0.74 & 11168 & 9258 & 1.858574055973438 \\
160519006001269 & 5186 & 68 & 3.152 & 0.109 & 0.103 & 0.056 & 1.404 & 2.204 & 1.15 & 9834 & 8094 & 1.8117238798172863 \\
210522002101019 & 5965 & 72 & 3.784 & 0.100 & 0.015 & 0.061 & 0.860 & 1.374 & 3.16 & 6642 & 6379 & 2.4510478480166302 \\
170414002601054 & 5906 & 71 & 3.826 & 0.120 & -0.225 & 0.062 & 0.721 & 1.183 & 4.81 & 6500 & 6417 & 1.4036048209441487 \\
140708000601266 & 6070 & 72 & 3.774 & 0.102 & -0.094 & 0.064 & 0.903 & 1.362 & 2.98 & 6818 & 6502 & 1.8280615721007407 \\
\enddata
\tablecomments{This table is published in its entirety in machine-readable format. A portion is shown here for guidance regarding its form and content. The full table contains 22,442 rows and 13 columns.}
\end{deluxetable*}

% ── Column key ──
\begin{deluxetable}{lll}
\tablecaption{Column descriptions for Table~\ref{tab:pms_undep}\label{tab:pms_undep_cols}}
\tablehead{
\colhead{Column} & \colhead{Label} & \colhead{Description}
}
\startdata
\texttt{galah\_id} & GALAH ID & GALAH source identifier \\
\texttt{teff} & $T_{\rm eff}$ & Effective temperature (K) \\
\texttt{teff\_err} & $\sigma_{T_{\rm eff}}$ & Uncertainty in $T_{\rm eff}$ (K) \\
\texttt{logg} & $\log g$ & Surface gravity (dex) \\
\texttt{logg\_err} & $\sigma_{\log g}$ & Uncertainty in $\log g$ (dex) \\
\texttt{feh} & [Fe/H] & Metallicity (dex) \\
\texttt{e\_fe\_h} & $\sigma_{\rm [Fe/H]}$ & Uncertainty in [Fe/H] (dex) \\
\texttt{loglbol} & $\log(L/L_\odot)$ & Bolometric luminosity \\
\texttt{mass} & $M$ & Stellar mass ($M_\odot$) \\
\texttt{age} & Age & Stellar age (Gyr) \\
\texttt{teff\_zams} & $T_{\rm eff,ZAMS}$ & YREC ZAMS $T_{\rm eff}$ (K) \\
\texttt{teff\_msto} & $T_{\rm eff,MSTO}$ & YREC MSTO $T_{\rm eff}$ (K) \\
\texttt{pms\_undep} & pms\_undep & Li corrrected for Pre-MS Depletion \\
\enddata
\end{deluxetable}

%% file: table3_print.tex
% 10-row excerpt
% Full table: 22,442 rows x 13 columns
% Print columns: 13

\begin{deluxetable*}{ccccccccccccc}
\tabletypesize{\scriptsize}
\tablecaption{Conservative Sample Stellar Parameters with YREC-Mapped Temperatures and RGB-Base Lithium Abundances\label{tab:rgb_ali}}
\tablecolumns{13}
\tablehead{
\colhead{GALAH ID} & \colhead{$T_{\rm eff}$} & \colhead{$\sigma_{T_{\rm eff}}$} & \colhead{$\log g$} & \colhead{$\sigma_{\log g}$} & \colhead{[Fe/H]} & \colhead{$\sigma_{\rm [Fe/H]}$} & \colhead{$\log(L/L_\odot)$} & \colhead{$M$} & \colhead{Age} & \colhead{$T_{\rm eff,ZAMS}$} & \colhead{$T_{\rm eff,MSTO}$} & \colhead{A(Li)$_{\rm RGB}$}
}
\startdata
160519002601288 & 5101 & 72 & 3.731 & 0.095 & -0.007 & 0.059 & 0.513 & 1.120 & 8.55 & 6039 & 6100 & -0.11 \\
150606001901019 & 5277 & 77 & 3.602 & 0.107 & -0.431 & 0.064 & 0.737 & 1.164 & 5.19 & 6665 & 6584 & 0.59 \\
170415002001297 & 5575 & 68 & 3.963 & 0.102 & 0.357 & 0.055 & 0.477 & 1.188 & 6.84 & 5758 & 5882 & -5.18 \\
150606004901094 & 5173 & 75 & 3.693 & 0.103 & -0.212 & 0.062 & 0.574 & 1.074 & 7.85 & 6173 & 6225 & 0.51 \\
170510007301313 & 6150 & 69 & 3.852 & 0.115 & -0.401 & 0.067 & 0.764 & 1.148 & 4.50 & 6614 & 6549 & 0.01 \\
230203002101025 & 5728 & 69 & 3.013 & 0.104 & -0.084 & 0.059 & 1.774 & 2.404 & 0.74 & 11168 & 9258 & 0.03 \\
160519006001269 & 5186 & 68 & 3.152 & 0.109 & 0.103 & 0.056 & 1.404 & 2.204 & 1.15 & 9834 & 8094 & 0.01 \\
210522002101019 & 5965 & 72 & 3.784 & 0.100 & 0.015 & 0.061 & 0.860 & 1.374 & 3.16 & 6642 & 6379 & 0.55 \\
170414002601054 & 5906 & 71 & 3.826 & 0.120 & -0.225 & 0.062 & 0.721 & 1.183 & 4.81 & 6500 & 6417 & -0.30 \\
140708000601266 & 6070 & 72 & 3.774 & 0.102 & -0.094 & 0.064 & 0.903 & 1.362 & 2.98 & 6818 & 6502 & 0.06 \\
\enddata
\tablecomments{This table is published in its entirety in machine-readable format. A portion is shown here for guidance regarding its form and content. The full table contains 22,442 rows and 13 columns.}
\end{deluxetable*}

% ── Column key ──
\begin{deluxetable}{lll}
\tablecaption{Column descriptions for Table~\ref{tab:rgb_ali}\label{tab:rgb_ali_cols}}
\tablehead{
\colhead{Column} & \colhead{Label} & \colhead{Description}
}
\startdata
\texttt{galah\_id} & GALAH ID & GALAH source identifier \\
\texttt{teff} & $T_{\rm eff}$ & Effective temperature (K) \\
\texttt{teff\_err} & $\sigma_{T_{\rm eff}}$ & Uncertainty in $T_{\rm eff}$ (K) \\
\texttt{logg} & $\log g$ & Surface gravity (dex) \\
\texttt{logg\_err} & $\sigma_{\log g}$ & Uncertainty in $\log g$ (dex) \\
\texttt{feh} & [Fe/H] & Metallicity (dex) \\
\texttt{e\_fe\_h} & $\sigma_{\rm [Fe/H]}$ & Uncertainty in [Fe/H] (dex) \\
\texttt{loglbol} & $\log(L/L_\odot)$ & Bolometric luminosity \\
\texttt{mass} & $M$ & Stellar mass ($M_\odot$) \\
\texttt{age} & Age & Stellar age (Gyr) \\
\texttt{teff\_zams} & $T_{\rm eff,ZAMS}$ & YREC ZAMS $T_{\rm eff}$ (K) \\
\texttt{teff\_msto} & $T_{\rm eff,MSTO}$ & YREC MSTO $T_{\rm eff}$ (K) \\
\texttt{rgb\_ali} & A(Li)$_{\rm RGB}$ & A(Li) at the base of the RGB \\
\enddata
\end{deluxetable}

%% file: table4_print.tex
% 10-row excerpt
% Full table: 22,442 rows x 15 columns
% Print columns: 15

\begin{deluxetable*}{ccccccccccccccc}
\tabletypesize{\scriptsize}
\tablecaption{Conservative Sample Lithium Abundances and First Dredge-Up Corrected Values\label{tab:ali_dredge}}
\tablecolumns{15}
\tablehead{
\colhead{GALAH ID} & \colhead{$T_{\rm eff}$} & \colhead{$\sigma_{T_{\rm eff}}$} & \colhead{$\log g$} & \colhead{$\sigma_{\log g}$} & \colhead{[Fe/H]} & \colhead{$\sigma_{\rm [Fe/H]}$} & \colhead{$\log(L/L_\odot)$} & \colhead{$M$} & \colhead{Age} & \colhead{$T_{\rm eff,ZAMS}$} & \colhead{$T_{\rm eff,MSTO}$} & \colhead{A(Li)} & \colhead{A(Li)$_{\rm up}$} & \colhead{A(Li)$_{\rm FDU}$}
}
\startdata
160519002601288 & 5101 & 72 & 3.731 & 0.095 & -0.007 & 0.059 & 0.513 & 1.120 & 8.55 & 6039 & 6100 & 1.00 & \nodata & 2.18 \\
150606001901019 & 5277 & 77 & 3.602 & 0.107 & -0.431 & 0.064 & 0.737 & 1.164 & 5.19 & 6665 & 6584 & \nodata & 0.91 & 2.20 \\
170415002001297 & 5575 & 68 & 3.963 & 0.102 & 0.357 & 0.055 & 0.477 & 1.188 & 6.84 & 5758 & 5882 & 1.98 & \nodata & 2.19 \\
150606004901094 & 5173 & 75 & 3.693 & 0.103 & -0.212 & 0.062 & 0.574 & 1.074 & 7.85 & 6173 & 6225 & 0.99 & \nodata & 2.19 \\
170510007301313 & 6150 & 69 & 3.852 & 0.115 & -0.401 & 0.067 & 0.764 & 1.148 & 4.50 & 6614 & 6549 & \nodata & 1.61 & 1.61 \\
230203002101025 & 5728 & 69 & 3.013 & 0.104 & -0.084 & 0.059 & 1.774 & 2.404 & 0.74 & 11168 & 9258 & 1.86 & \nodata & 1.86 \\
160519006001269 & 5186 & 68 & 3.152 & 0.109 & 0.103 & 0.056 & 1.404 & 2.204 & 1.15 & 9834 & 8094 & \nodata & 0.53 & 1.81 \\
210522002101019 & 5965 & 72 & 3.784 & 0.100 & 0.015 & 0.061 & 0.860 & 1.374 & 3.16 & 6642 & 6379 & 2.41 & \nodata & 2.41 \\
170414002601054 & 5906 & 71 & 3.826 & 0.120 & -0.225 & 0.062 & 0.721 & 1.183 & 4.81 & 6500 & 6417 & \nodata & 1.37 & 1.38 \\
140708000601266 & 6070 & 72 & 3.774 & 0.102 & -0.094 & 0.064 & 0.903 & 1.362 & 2.98 & 6818 & 6502 & 1.81 & \nodata & 1.81 \\
\enddata
\tablecomments{This table is published in its entirety in machine-readable format. A portion is shown here for guidance regarding its form and content. The full table contains 22,442 rows and 15 columns.}
\end{deluxetable*}

% ── Column key ──
\begin{deluxetable}{lll}
\tablecaption{Column descriptions for Table~\ref{tab:ali_dredge}\label{tab:ali_dredge_cols}}
\tablehead{
\colhead{Column} & \colhead{Label} & \colhead{Description}
}
\startdata
\texttt{galah\_id} & GALAH ID & GALAH source identifier \\
\texttt{teff} & $T_{\rm eff}$ & Effective temperature (K) \\
\texttt{teff\_err} & $\sigma_{T_{\rm eff}}$ & Uncertainty in $T_{\rm eff}$ (K) \\
\texttt{logg} & $\log g$ & Surface gravity (dex) \\
\texttt{logg\_err} & $\sigma_{\log g}$ & Uncertainty in $\log g$ (dex) \\
\texttt{feh} & [Fe/H] & Metallicity (dex) \\
\texttt{e\_fe\_h} & $\sigma_{\rm [Fe/H]}$ & Uncertainty in [Fe/H] (dex) \\
\texttt{loglbol} & $\log(L/L_\odot)$ & Bolometric luminosity \\
\texttt{mass} & $M$ & Stellar mass ($M_\odot$) \\
\texttt{age} & Age & Stellar age (Gyr) \\
\texttt{teff\_zams} & $T_{\rm eff,ZAMS}$ & YREC ZAMS $T_{\rm eff}$ (K) \\
\texttt{teff\_msto} & $T_{\rm eff,MSTO}$ & YREC MSTO $T_{\rm eff}$ (K) \\
\texttt{ali} & A(Li) & Lithium abundance (3D NLTE) \\
\texttt{ali\_upper} & A(Li)$_{\rm up}$ & Upper limit on A(Li) \\
\texttt{dredge} & A(Li)$_{\rm FDU}$ & FDU-corrected A(Li) \\
\enddata
\end{deluxetable}

%% file: table5_print.tex
% 10-row excerpt
% Full table: 22,442 rows x 3 columns
% Print columns: 3

\begin{deluxetable*}{ccc}
\tablecaption{Conservative Sample Cross-Matched Identifiers: GALAH, Gaia DR3, and 2MASS\label{tab:ids}}
\tablecolumns{3}
\tablehead{
\colhead{GALAH ID} & \colhead{Gaia DR3 ID} & \colhead{2MASS ID}
}
\startdata
160519002601288 & 5377030709335263616 & 11175717-4346408 \\
150606001901019 & 5387232184375415936 & 11023509-4500146 \\
170415002001297 & 5742287883667654656 & 09142630-1049156 \\
150606004901094 & 6713562306363299584 & 19193234-4135329 \\
170510007301313 & 6664918537561478784 & 19293196-4351586 \\
230203002101025 & 3040524794725372928 & 07462162-1021061 \\
160519006001269 & 6725162909979376384 & 18055081-4216514 \\
210522002101019 & 5840608554194329088 & 13321600-7213321 \\
170414002601054 & 5467140944268942720 & 10065358-2549044 \\
140708000601266 & 6096274007263370368 & 14150972-4551029 \\
\enddata
\tablecomments{This table is published in its entirety in machine-readable format. A portion is shown here for guidance regarding its form and content. The full table contains 22,442 rows and 3 columns.}
\end{deluxetable*}

% ── Column key ──
\begin{deluxetable}{lll}
\tablecaption{Column descriptions for Table~\ref{tab:ids}\label{tab:ids_cols}}
\tablehead{
\colhead{Column} & \colhead{Label} & \colhead{Description}
}
\startdata
\texttt{galah\_id} & GALAH ID & GALAH source identifier \\
\texttt{gaiadr3\_source\_id} & Gaia DR3 ID & Gaia DR3 source identifier \\
\texttt{tmass\_id} & 2MASS ID & 2MASS source identifier \\
\enddata
\end{deluxetable}